\documentclass[
preprint, 
preprintnumbers,superscriptaddress,showpacs,supberscriptaddress,
nofootinbib,longbibliography]{revtex4-1}

\usepackage[
bookmarks=true,colorlinks,linkcolor=blue,urlcolor=cyan,citecolor=red]{hyperref}

\usepackage{graphicx}
\usepackage{latexsym}
\usepackage{amsfonts}
\usepackage{amssymb}
\usepackage{amsmath}
\usepackage{bm}
\usepackage{mathrsfs}
\usepackage{slashed}
\usepackage{ascmac}

\usepackage{dcolumn}

\allowdisplaybreaks[3]

\newcommand{\nn}{\nonumber}

\newcommand{\sla}{ \slashed }

\newcommand{\ep}{ \epsilon }

\newcommand{\gam}{ \gamma }

\newcommand{\pint}{\int \frac{d^4p}{ (2\pi)^4}}

\newcommand{\id}{\mbox{1}\hspace{-0.25em}\mbox{l}}

\newcommand{\nonA}{ {\mathcal A} }
\newcommand{\nonF}{ {\mathcal{F}} }
\newcommand{\nonB}{ {\mathcal{B}} }

\newcommand{\F}{ {\mathscr{F}} }
\newcommand{\G}{ {\mathscr{G}} }

\renewcommand{\Finv}{ {\mathscr{F}} }
\newcommand{\Ginv}{ {\mathscr{G}} }

\newcommand{\bB}{{\bm B}}
\newcommand{\bE}{{\bm E}}

\newcommand{\ext}{ {\rm ext} }

\newcommand{\tr}{ {\rm tr} }
\newcommand{\diag}{ {\rm diag} }
\newcommand{\sgn}{ {\rm sgn} }

\newcommand{\Lag}{ {\cal{L}} }

\newcommand{\quark}{ {\rm quark} }
\newcommand{\gluon}{ {\rm gluon} }
\newcommand{\ghost}{ {\rm ghost} }

\newcommand{\para}{ \parallel}

\newcommand{\J}{ {\mathcal J} }

\newcommand{\del}{\partial}
\usepackage{comment}
\usepackage{epstopdf}

\newcommand{\e}{{\rm e}}
\newcommand{\aaa}{\mathfrak{a}}
\newcommand{\bbb}{\mathfrak{b}}

\newcommand{\order}{ {{\mathcal O} }}

\newcommand{\beq}{\begin{eqnarray}}
\newcommand{\eeq}{\end{eqnarray}}
\newcommand{\bseq}{\begin{subequations}}
\newcommand{\eseq}{\end{subequations}}

\usepackage{color} 
\usepackage[normalem]{ulem}  
\newcommand{\com}[1]{{\color[rgb]{0,0,1}{#1}}}

\begin{document}

\title{Note on all-order Landau-level structures of 
\\
the Heisenberg-Euler effective actions for QED and QCD}

\author{Koichi Hattori}
\affiliation{Yukawa Institute for Theoretical Physics, Kyoto University, Kyoto 606-8502, Japan.}

\author{Kazunori Itakura}
\affiliation{KEK Theory Center, Institute of Particle and Nuclear Studies,
High Energy Accelerator Research Organization, 1-1, Oho, Ibaraki, 305-0801, Japan.}
\affiliation{Graduate University for Advanced Studies (SOKENDAI), 
1-1 Oho, Tsukuba, Ibaraki 305-0801, Japan.}

\author{Sho Ozaki}
\affiliation{Department of Radiology, University of Tokyo Hospital,
7-3-1, Hongo, Bunkyo-ku, Tokyo 113-8655, Japan.}

\preprint{YITP-20-03}

\begin{abstract}
We investigate the Landau-level structures encoded in 
the famous Heisenberg-Euler (HE) effective action in constant electromagnetic fields. 
We first discuss the HE effective actions for scalar and spinor QED, 
and then extend it to the QCD analogue in the covariantly constant chromo-electromagnetic fields. 
We identify all the Landau levels and the Zeeman energies starting out 
from the proper-time representations at the one-loop order, 
and derive the vacuum persistence probability for the Schwinger mechanism 
in the summation form over independent contributions of the all-order Landau levels. 
We find an enhancement of the Schwinger mechanism {\it catalyzed} by a magnetic field for spinor QED 
and, in contrast, a stronger exponential suppression for scalar QED 
due to the ``zero-point energy'' of the Landau quantization. 
For QCD, we identify the discretized energy levels of the transverse and longitudinal gluon modes 
on the basis of their distinct Zeeman energies, 
and explicitly confirm the cancellation between the longitudinal-gluon and ghost contributions in the Schwinger mechanism. 
We also discuss the unstable ground state of the perturbative gluon excitations 
known as the Nielsen-Olesen instability.

\end{abstract}

\maketitle

\section{Introduction}

Heisenberg and Euler opened a new avenue toward the strong-field QED 
with their famous low-energy effective theory \cite{Heisenberg:1935qt} 
many years ahead of systematic understanding of QED. 
Some time later, Schwinger reformulated the Heisenberg-Euler (HE) effective action by 
the use of the proper-time method \cite{Schwinger:1951nm} 
which was discussed by Nambu \cite{Nambu:1950rs} and Feynman \cite{Feynman:1950ir} 
on the basis of the idea of introducing the proper time as an independent parameter of the motion by Fock \cite{Fock:1937dy}. 
Since then, the HE effective action has been playing the central role 
on describing the fundamental quantum dynamics in the low-energy, but intense, electromagnetic fields. 
Especially, the HE effective action has been used to describe 
the pair production in a strong electric field \cite{Sauter:1931zz, Heisenberg:1935qt, Schwinger:1951nm} 
and the effective interactions among the low-energy photons that give rise to the nonlinear QED effects 
such as the vacuum birefringence and photon splitting \cite{Toll:1952rq, Klein:1964zza, Baier:1967zza, Baier:1967zzc,BialynickaBirula:1970vy,Brezin:1971nd,Adler:1971wn} (see Ref.~\cite{Dittrich:2000zu} for a review article).

The HE effective action was also extended to its non-Abelian analogue in the chromo-electromagnetic field. 
The prominent difference in non-Abelian theories from QED is the presence of 
the self-interactions among the gauge bosons. 
The contribution of the gauge-boson loop provides the logarithmic singularity 
at the vanishing chromo-magnetic field limit \cite{Batalin:1976uv}, 
inducing a non-trivial minimum of the effective potential at a finite value of the chromo-magnetic field. 
This suggests the formation of a coherent chromo-magnetic field, 
or the ``magnetic gluon condensation'', in the QCD vacuum 
and the logarithmic singularity was also shown to reproduce the negative beta function of QCD \cite{Batalin:1976uv, Matinyan:1976mp, Savvidy:1977as, Yildiz:1979vv, Dittrich:1983ej} (see Ref.~\cite{Savvidy:2019grj} for a recent retrospective review paper). 
About a half of the logarithm comes from the tachyonic ground state of the gluon spectrum 
known as the Nielsen-Olesen unstable mode \cite{Nielsen:1978rm} 
that is subject to the Landau quantization and the negative Zeeman shift  in the chromo-magnetic field. 
On the other hand, the quark-loop contribution in the chromo-electromagnetic field was applied to 
the quark and antiquark pair production in the color flux tubes \cite{Casher:1978wy, Casher:1979gw} 
and the particle production mechanism in the relativistic heavy-ion collsions \cite{Biro:1984cf, Kajantie:1985jh, Gyulassy:1986jq} (see Ref.~\cite{Gelis:2015kya} for a recent review paper). 
The quark-loop contribution was more recently generalized to the case under 
the coexisting chromo and Abelian electromagnetic fields \cite{Ozaki:2013sfa, Bali:2013esa}.

The HE effective action has been also the main building block to describe the chiral symmetry breaking 
in QED and QCD under strong magnetic fields (see, e.g., Refs.~\cite{Dittrich:1978fc, Klevansky:1989vi, Suganuma:1990nn, 
Gusynin:1994xp, Gusynin:1995nb, Cohen:2007bt, Mizher:2010zb, Skokov:2011ib, Fukushima:2012xw} 
and a review article \cite{Andersen:2014xxa}). 
Note that the chiral symmetry breaking occurs even in weak-coupling theories 
in the strong magnetic field \cite{Gusynin:1995gt}. 
Furthermore, the HE effective action in the chromo-field, ``$ A^0 $ background'', at finite temperatures reproduces 
the Weiss-Gross-Pisarski-Yaffe potential \cite{Gross:1980br, Weiss:1980rj} for the Polyakov loop \cite{Gies:2000dw}. 
Therefore, it could be generalized to the case under the influence of the coexisting Abelian magnetic field \cite{Ozaki:2015yja}. 
Those analytic results can be now compared with the lattice QCD studies (see the most recent results \cite{DElia:2018xwo, Tomiya:2019nym, Endrodi:2019zrl} and references therein for the numerical efforts and novel observations over the decade).

The HE effective action is one of the fundamental quantities which have a wide spectrum of applications. 
While the proper-time method allows for the famous representation of the HE effective action in a compact form, 
it somewhat obscures the physical content encoded in the theory. 
Therefore, in this note, we clarify the Landau-level structure of the HE effective action by analytic methods. 
While two of the present authors showed the analytic structure of the one-loop vacuum polarization diagram, 
or the two-point function, in Refs.~\cite{Hattori:2012je, Hattori:2012ny} (see also Ref.~\cite{Ishikawa:2013fxa}),  
we find a much simpler form for the HE effective action, the zero-point function. 
As an application of the general formula, we compute the imaginary part of the effective action 
which explicitly indicates the occurrence of the Schwinger mechanism with an infinite sequence of 
the critical electric fields defined with the Landau levels. 
Those critical fields may be called the Landau-Schwinger limits. 
We here maintain the most general covariant form of the constant electromagnetic field configurations, 
expressed with the Poincar\`e invariants, 
and discuss qualitative differences among the effects of the magnetic field on the Schwinger mechanism 
in different field configurations. 
The parallel electromagnetic field configuration was recently discussed 
with the Wigner-function formalism \cite{Sheng:2018jwf}.\footnote{
We thank Shu Lin for drawing our attention to this reference.
}

In Sec.~\ref{sec:PTM}, we summarize the derivation of the HE effective action by the proper-time method. 
Then, we discuss the Landau-level structures of the HE effective actions in Sec.~\ref{sec:QED} for QED, 
and in Sec.~\ref{sec:QCD} for QCD in the covariantly constant field. 
In appendices, we supplement some technical details. 
We use the mostly minus signature of the Minkowski metric $g^{\mu\nu} = {\rm diag}(1,-1,-1,-1)$ 
and the completely antisymmetric tensor with $ \ep^{0123} =+ 1 $.

\section{Effective actions in scalar and spinor QED}

\label{sec:PTM}

In this section, we 
present a careful 
derivation of the HE effective action 
by the proper-time method {\it a la} Schwinger~\cite{Schwinger:1951nm}. 
This formalism is also useful to investigate the Landau-level structures in the forthcoming sections.

\subsection{Proper-time method}

\begin{figure}[t]
\begin{center}
   \includegraphics[width=0.9\hsize]{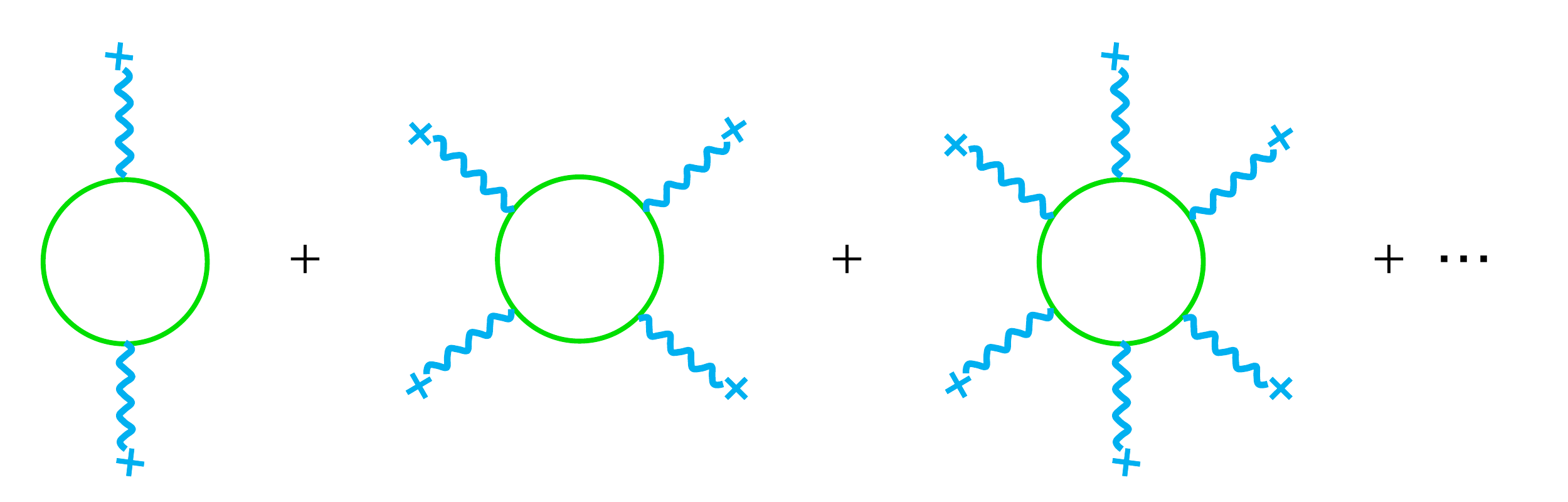}
\end{center}
\caption{
One-loop diagrams contributing to the Heisenberg-Euler effective action 
which is obtained by integrating out the matter field (green solid lines). 
The diagrams with odd-number insertions of the electromagnetic field vanish 
due to Furry's theorem. 
}
  \label{fig:OneLoopHE}
\end{figure}

We first introduce the proper-time method for scalar QED and then proceed to spinor QED. 
The classical Lagrangian of scalar QED is given by 
\begin{eqnarray}
\Lag_s = (D_\mu \phi)^\ast (D^\mu \phi) - m^2 \phi^\ast \phi
\label{eq:Lag-s}
\, .
\end{eqnarray}
Our convention of the covariant derivative is $D_\mu=\del_\mu+iq_f A_\mu $, 
where the electrical charge $q_f$ is negative for, e.g., electrons ($q_f=-|e|$). 
The gauge field $ A_\mu $ is for the external field, and we do not consider the dynamical gauge field. 
This corresponds to the one-loop approximation for the effective action. 
We write the classical and one-loop contribution to the Lagrangian as ${\cal L}_{\rm eff}={\cal L}^{(0)}+{\cal L}^{(1)}$ 
with the Maxwell term $ {\cal L}^{(0)} = - F_{\mu\nu}F^{\mu\nu}/4$ and the one-loop correction ${\cal L}^{(1)} $. 
The effective action $S_{\rm eff}=\int d^4x\, {\cal L}_{\rm eff}[A_\mu]$ is formally obtained 
by the path-integration of the classical action with respect to the bilinear matter field. 
In case of scalar QED, we find the determinant of the Klein-Gordon operator 
\beq
 S^{(1)} _s [A_\mu]=-i\ln \det (D^2 + m^2)^{-\frac{2}{2}} = i \ln \det (D^2 + m^2)
\, .
\label{Lag_original-s}
\eeq 
The determinant is doubled for the two degrees of freedom in the complex scalar field. 
Diagrammatically, the quantum correction ${\cal L}^{(1)}$ corresponds to the one-loop contributions in Fig.~\ref{fig:OneLoopHE} 
which are 
summed with respect to the external-field insertions to the infinite order.

We shall introduce a useful formalism called the proper-time method, 
which was discussed by Nambu \cite{Nambu:1950rs} and Feynman \cite{Feynman:1950ir} 
on the basis of the idea of introducing the proper time as an independent parameter of the motion by Fock \cite{Fock:1937dy} 
and was finally established by Schwinger \cite{Schwinger:1951nm}. 
By the use of a formula\footnote{
One can obtain this formula 
by integrating the both sides of the identify 
$  (X - i \epsilon)^{-1} = i \int_0^\infty ds \, \e^{-is(X-i\epsilon)}$, 
with respect to $  X$ from $ B $ to $A  $. 
}
\begin{eqnarray}
\ln\frac{A-i\epsilon}{B-i\epsilon} = - \int_0^\infty \frac{ds}{s} \left({\rm e}^{ - is(A - i\epsilon)} - {\rm e}^{-is(B - i\epsilon)}\right)
\label{propertime2}
\, ,
\end{eqnarray}
one can rewrite the one-loop correction in the integral form  
\beq
{\cal L}_s^{(1')}= - i \int_0^\infty \frac{ds}{s}\, {\rm e}^{-is(m-i\epsilon)^2} 
\left[ 
\langle x | {\rm e}^{-i \hat H_s s}|x\rangle - \langle x|{\rm e}^{-i \hat H_{s0} s}|x\rangle 
\right] 
\label{effectiveaction-s}
\, .
\eeq
We applied a familiar formula $ \ln \det \order  = \tr \ln \order $, 
and took the trace over the coordinate space. 
An infinitesimal positive parameter $\epsilon>0$ ensures the convergence of the integral with respect to $ s $. 
The integral variable $  s$ is called the proper time since it is parametrizing 
the proper-time evolution governed by the ``Hamiltonian" 
\beq
\hat H_s \equiv D^2 
\label{Ham_scalar}
\, .
\eeq 
In Eq.~(\ref{effectiveaction-s}), 
we have subtracted the free-theory contribution in the absence of external fields 
evolving with the free Hamiltonian $\hat H_{s0}=\del^2$. 
The Lagrangian (\ref{effectiveaction-s}) is marked with a prime after the subtraction. 
We may also define the ``time-evolution operator'' 
\begin{eqnarray}
 \hat U(x;s) \equiv  {\rm e}^{-i \hat H_s s}
 \, .
\end{eqnarray}
An advantage of the proper-time method is that the quantum field theory problem 
has reduced to a quantum mechanical one. 
We will solve the counterparts of the Schr\"odinger and Heisenberg equations for the proper-time evolution.

Before solving the problem, we summarize a difference between spinor and scalar QED. 
We apply the proper-time method to spinor QED of which the classical Lagrangian is given as\footnote{
We do not explicitly distinguish the mass parameters of the scalar particle and the fermion 
since the coupling among those fields is not considered in this paper. 
} 
\begin{eqnarray}
{\cal L}_f = \bar \psi (i \sla D - m) \psi
\, .
\end{eqnarray}
Performing the path integration over the fermion bilinear field, 
the effective action is given by the determinant of the Dirac operator 
\beq
S^{(1)}_f [A_\mu]=-i\ln \det (i\sla D-m)=-\frac{i}{2}\ln \det (\sla D^2 + m^2)
\label{Lag_original-f}
\, .
\eeq 
To reach the last expression, we have used a relation $\det (i \sla D-m)=\det (i \sla D+m)$ 
which holds thanks to the charge conjugation symmetry $C \sla D C^{-1}=- \sla D ^T$. 
Again, we can rewrite Eq.~(\ref{Lag_original-f}) by using the formula~(\ref{propertime2}) with the proper time $s$: 
\beq
{\cal L}_f^{(1')}=\frac{i}{2}\int_0^\infty \frac{ds}{s}\,  {\rm e}^{-is(m-i\epsilon)^2} 
{\rm tr}\, \left[ 
\langle x | {\rm e}^{-i \hat H_f s}|x\rangle - \langle x|{\rm e}^{-i \hat H_{f0} s}|x\rangle 
\right] 
\, ,
\label{effectiveaction-f}
\eeq
where  ``tr" indicates the trace over the Dirac spinor indices. 
The ``Hamiltonian" is defined as 
\beq
\hat H_f \equiv D^2 + \frac{q_f}{2}F^{\mu\nu}\sigma_{\mu \nu}
\label{Ham_fermion}
\, ,
\eeq 
where $ \sigma^{\mu\nu} = \frac{i}{2}  [ \gam^\mu, \gam^\nu] $ 
and $ \hat H_{f0}  =\del^2\id \ (=   \hat H_{s0} \id)$ with the unit matrix $ \id $ in the spinor space.

Accordingly, the difference between the scalar and spinor QED is found to be 
\beq
 \hat H_f - \hat H_s \id  = \frac{q_f}{2}F^{\mu\nu}\sigma_{\mu \nu}
\, .
\eeq
The difference originates from the spinor structure in the squared Dirac operator $ \sla D^2 $, 
and thus is responsible for the spin interaction with the external field. 
The scalar term $ D^2 $ and the spin-interaction term commute with each other 
when the external field $ F^{\mu\nu} $ is constant, 
so that the spin-interaction term can be factorized as a separate exponential factor.  
As shown in Appendix~\ref{sec:traces}, the trace of the spin part can be carried out as 
\begin{eqnarray}
\label{eq:F-sigma}
\tr \left[ \e^{- is  \frac{q_f}{2}F^{\mu\nu}\sigma_{\mu \nu}} \right] 
= 4 \cosh( q_f s a ) \cos(q_f sb) 
\, .
\end{eqnarray}
Then, we find the relation between the transition amplitudes in scalar and spinor QED 
\begin{eqnarray}
\label{eq:s-f}
\tr \langle x | {\rm e}^{-i \hat H_f s}|x\rangle 
=  \left[ 4 \cosh( q_f s a ) \cos( q_f sb)  \right]  \times \langle x | {\rm e}^{-i  \hat H_s s}|x\rangle 
\, .
\end{eqnarray}
Therefore, in the next section, 
we can focus on the scalar transition amplitude, $  \langle x | {\rm e}^{-i  \hat H_s s}|x\rangle  $.

\subsection{Coordinate representation}

We first need to provide a set of boundary conditions to solve the equation of motion. 
We consider a transition from $ x_0^\mu $ to $ x_1^\mu $ when the proper time evolves from $ 0 $ to $ s $, 
and the coincidence limit  $ x_1^\mu \to x_0^\mu $ (with a finite value of $  s$ maintained) 
that is necessary for the computation of the transition amplitude $  \langle x | {\rm e}^{-i  \hat H_s s}|x\rangle  $ 
in the HE effective action. 


In the free theory, one can immediately find the transition amplitude 
\begin{eqnarray}
\label{eq:W0}
 \langle x_1 | {\rm e}^{-i  \hat H_{s0} s}|x_0 \rangle = \pint  \e^{-ip(x_1-x_0)} {\rm e}^{ ip^2s} 
   =- \frac{i}{(4\pi)^2 s^2 } \e^{  - \frac{i}{4s}   (x_1-x_0)^2 }
 \, .
\end{eqnarray}
Unlike quantum mechanics, the transition amplitude does not reduce to unity when $ s \to 0 $, but actually diverges. 
This is a manifestation of the ultraviolet singularity in quantum field theory. 
The coincidence limit is obtained as 
\begin{eqnarray}
\lim_{x_1\to x}  \langle x_1 | {\rm e}^{-i  \hat H_{s0} s}|x \rangle  
 = - \frac{i}{(4\pi)^2 s^2} 
 \, .
\end{eqnarray}

In the presence of an external field, we need to solve the ``Shr\"odinger equation'' 
\begin{eqnarray}
\label{eq:Schrodinger}
i \frac{ d W(x;s)}{ ds} = \langle x_1 | \hat U(x;s) \hat H_s  | x_0 \rangle
= \langle x(s) |  \hat H_s  | x (0)\rangle
\, ,
\end{eqnarray}
where the ``transition amplitude'' is defined as 
$  W(x;s): =  \langle x_1 |  \hat U(x;s) | x_0 \rangle  =  \langle x (s) |  x (0) \rangle  $. 
In the Heisenberg picture, the basis evolves as $ | x (s) \rangle =  \hat U^\dagger(x;s) | x_1 \rangle $, 
while the state is intact, $  | x_0 \rangle =  | x(0) \rangle $.  
The Heisenberg equations for the operators $\hat x^\mu(s)  $ and $ \hat D^\mu(s) $ are given as 
\begin{subequations}
\label{eq:Heisenberg}
\begin{eqnarray}
\label{eq:Heisenberg-x}
&&
\frac{d \hat x^\mu(s)}{ds} = i[ \hat H_s  ,\hat  x^\mu(s) ] = 2 i \hat  D^\mu(s)
\, ,
\\
\label{eq:Heisenberg-D}
&&
\frac{d \hat D_\mu(s)}{ds} = i[ \hat H_s  , \hat D_\mu(s) ]  = 2 q_f F_{\mu}^{\ \, \nu} \hat  D_\nu(s)
\, .
\end{eqnarray}
\end{subequations}
The second equation holds for constant field strength tensors. 
The solutions of those equations are straightforwardly obtained as 
\begin{subequations}
\begin{eqnarray}
&&
\hat D_\mu(s) = \e^{2q_f s F_{\mu}^{\  \nu}} \hat D_\nu (0)
\, ,
\\
&&
\hat x_\mu(s) -\hat  x_\mu(0) = 
i q_f^{-1} (F^{-1})_{\mu}^{\ \,  \nu} ( \e^{2q_f s F_{\nu}^{\  \sigma}} - \delta_{\nu}^{\sigma}) \hat D_\sigma(0)
\, ,
\end{eqnarray}
\end{subequations}
where $  (F^{-1})_{\mu}^{\ \, \nu} $ is the inverse matrix of $  F_{\mu}^{\ \, \nu} $. 
Since they can be interpreted as matrices, 
we hereafter write them and other vectors without the Lorentz indices for the notational simplicity. 
Combining those two solutions, we get 
\begin{subequations}
\label{eq:D}
\begin{eqnarray}
\hat D(0) &=& \frac{1}{2i}  \sinh^{-1} (q_f F s)  \e^{ - q_f F s} (q_f F) [\hat x(s) -\hat  x(0) ]
\, ,
\\
\hat D(s) &=& \frac{1}{2i} \sinh^{-1} (q_f F s)  \e^{ q_f F s} (q_f F) [\hat x(s) -\hat  x(0) ]
\\
&=&  \frac{1}{2i} [\hat x(s) -\hat  x(0) ] (q_f F) \e^{-  q_f F s} \sinh^{-1} (q_f F s)
\, .
\end{eqnarray}
\end{subequations}
We have taken the transpose of the antisymmetric matrix in the last expression. 
Plugging those solutions into the Hamiltonian (\ref{Ham_scalar}), 
we have 
\begin{eqnarray}
\label{eq:Ham-bilinear}
\hat H_s 
=  [ \hat  x(s) - \hat x(0) ]  K(F,s) [\hat x(s) - \hat x(0) ] 
\, ,
\end{eqnarray}
with $ K (F,s):= (q_f F)^2 /[ (2i)^2 \sinh^{2} (q_f F s)] $. 
The vanishing field limit is $ K(0,s) = - 1/(4s^2) $. 
Since $ \hat x(s) $ contains $ \hat D(0) $, it does not commute with $\hat  x(0) $ but obeys a commutation relation 
\begin{eqnarray}
\label{eq:commutator}
[ \hat x(s), \hat x(0)] =2 i (q_f F)^{-1} \e^{  q_f F s} \sinh (q_f F s)
\, .
\end{eqnarray}
By using this commutator, we have 
\begin{eqnarray}
\hat H_s = \hat x(s) K \hat x(s) + \hat  x(0) K\hat  x(0) - 2\hat   x(s) K \hat x(0) + \frac{1}{ 2 i}  \tr[ (q_f F)  \coth (q_f F s)]
\, .
\end{eqnarray}
We have used an identify $   \tr[ (q_f F) \e^{  q_f F s} \sinh^{-1} (q_f F s)] =  \tr[  (q_f F)   \coth (q_f F s)] $, 
which follows from the fact that the trace of the odd-power terms vanishes, i.e., $ \tr [ F^{2n+1}]=0 $. 

Then, the coordinate representation of the Schr\"odinger equation (\ref{eq:Schrodinger}) reads 
\begin{eqnarray}
\label{eq:dK-x}
i \frac{ d W(x;s) }{ds} = \left[ \, ( x_1 - x_0  )  K ( x_1 - x_0 ) + \frac{1}{ 2 i}  \tr[ (q_f F)  \coth (q_f F s)]  \, \right] W(x;s) 
\, .
\end{eqnarray}
The solution is obtained in the exponential form 
\begin{eqnarray}
\label{eq:sol-x}
W(x;s) &=&  C_A (x_1,x_0)  \exp \left[  \, 
- \frac{i}{4} ( x_1 - x_0  ) (q_f F) \coth(q_f Fs) ( x_1 - x_0 ) 
 - \frac{1}{ 2 }  \tr[  \ln \{ \sinh (q_f F s) \} ]  
 \, \right]
 \nn
 \\
& \to&  C_A (x_1 \to x_0)  \exp \left[  - \frac{1}{ 2 }  \tr[  \ln \{ \sinh (q_f F s) \} ]   \, \right]
 \, .
\end{eqnarray}
The second line shows the coincidence limit 
which we need for the computation of the HE effective action 
and originates from the commutation relation (\ref{eq:commutator}). 
We could have an overall factor of $ C_A(x_1,x_0) $ as long as it is independent of $  s$. 
It is clear from Eq.~(\ref{eq:esinh}) that we should have the following factor of $ C_0(a,b) $ 
so that $ W(x;s) $ reduces to the free result (\ref{eq:W0}) in the vanishing field limit $ a,b \to0 $. 
Comparing those cases, we find 
\begin{eqnarray}
C_A(x_1,x_0) \propto C_0(a,b) = \frac{ (q_fa)(q_fb) }{ (4\pi)^2 }
\, .
\end{eqnarray}
Still, the $ C_A(x_1,x_0)  $ could have such multiplicative factors 
that depend on the external field but reduce to unity in the vanishing field limit. 
%
%
The residual part of $ C_A(x_1,x_0)  $ is determined by the normalization of the covariant derivative. 
One can evaluate the expectation value $ \langle x(s) | D^\mu(0)  | x(0) \rangle  $ in two ways by using Eq.~(\ref{eq:D}) 
and equate them to find the following equation in the coincidence limit ($ x_1^\mu \to x_0^\mu $): 
\begin{subequations}
\begin{eqnarray}
 &&
  (\partial_{x_0} + i q_f A(x_0)) \bar C_A (x_1 \to x_0) = 0
   \, ,
 \\
 &&
 (\partial_{x_1}^\mu + i q_f A(x_1)) \bar C_A^\dagger (x_1 \to x_0) = 0
  \, ,
\end{eqnarray}
\end{subequations}
where the second equation is obtained from $ \langle x(0) | D^\mu(s)  | x(s) \rangle  $ likewise. 
Therefore, we have 
\begin{eqnarray}
\label{eq:CA}
\bar  C_A (x_1 \to x_0) = C_0(a,b) \lim_{x_1 \to x_0} \exp \left[  i q_f \int_{x_0}^{x_1} dx^\mu A_\mu (x) \right] 
=  C_0(a,b)
\, .
\end{eqnarray}
The closed-contour integral vanishes unless there exists a non-trivial homotopy.

The remaining task is to compute the trace in the exponential factor. 
Note that the matrix form of the field strength tensor $  F_\mu^{\ \, \nu}$ satisfies 
an eigenvalue equation $ F_\mu^{\ \, \nu} \phi_\nu = \lambda \phi_\mu  $. 
The four eigenvalues are given by the Poincar\'e invariants $  \lambda = \pm a, \ \pm ib$ 
that are defined as
\begin{eqnarray}
\label{eq:Poincare}
 a= ( \sqrt{ \F^2 + \G^2 } - \F )^{1/2} , \quad b = ( \sqrt{ \F^2 + \G^2 } + \F )^{1/2}
 \, ,
\end{eqnarray}
with $\Finv \equiv  F_{\mu\nu} F^{\mu\nu}/4  $ 
and $\Ginv \equiv \epsilon^{\mu\nu\alpha\beta} F_{\mu\nu} F_{\alpha\beta} / 8 $. 
Therefore, we can immediately decompose the matrix into a simple form
\begin{eqnarray}
\label{eq:esinh}
\e^{ - \frac{1}{ 2 }  \tr[  \ln \{ \sinh (q_f F s) \} ] } 
= {\rm det} [ \sinh(q_f Fs)  ]^{-\frac{1}{2}} 
= \frac{ -i }{  \sinh (q_f a s) \sin (q_f b s)  }
\, .
\end{eqnarray}
Plugging this expression into 
Eqs.~(\ref{effectiveaction-s}) and (\ref{effectiveaction-f}) [see also Eq.~(\ref{eq:s-f})], 
we obtain the HE effective Lagrangians 
\begin{subequations}
\label{effectiveaction-x}
\begin{eqnarray}
{\cal L}_s^{(1')} &=& -
\frac{1}{16\pi^2}\int_0^\infty \frac{ds}{s}\,  {\rm e}^{-is(m^2-i\epsilon)} 
\left[\frac{ (q_f a )(q_f b ) }{\sinh (q_f a s) \sin (q_f b s)} - \frac{1}{s^2} \right] 
\, ,
\\
{\cal L}_f^{(1')} &=& 
 \frac{1}{8\pi^2}\int_0^\infty \frac{ds}{s}\,  {\rm e}^{-is(m^2-i\epsilon)} 
\left[ \frac{  (q_f a )(q_f b ) }{  \tanh( q_f s a ) \tan(q_f sb) } - \frac{1}{s^2} \right] 
\, ,
\end{eqnarray}
\end{subequations}
where the first and second ones are for scalar and spinor QED, respectively. 
Those results are manifestly gauge invariant 
after the phase factor vanishes in the coincidence limit (\ref{eq:CA}). 

In the series of the one-loop diagrams (cf. Fig.~\ref{fig:OneLoopHE}), 
the first and second diagrams with zero and two insertions of the external field give rise to ultraviolet divergences 
which appear in the small-$ s $ expansion of the integrand in the proper time method. 
The free-theory term has a divergence as noted earlier and works as one of the subtraction terms. 
Another divergence is easily identified in the small-$  s$ expansion of the integrand 
as $ \int \! ds\  q_f^2 (b^2-a^2)/(6s) $ and $ - \int \! ds \ q_f^2 (b^2-a^2)/(3s) $ 
(shown up to the overall factors) for scalar and spinor QED, respectively, 
and need to be subtracted for a complete renormalization procedure. 
The former and latter divergences should be dealt with the charge and field-strength renormalizations.

\section{All-order Landau-level structures}

\label{sec:QED} 

Having looked back the standard representation of the HE effective Lagrangian in the previous section, 
we now proceed to investigating the all-order Landau-level structures encoded in the effective Lagrangian.

\subsection{Momentum representation}


While we worked in the coordinate space in the previous section, 
we will take a slight detour via the momentum-space representation. 
After the gauge-dependent and translation-breaking phase 
has gone in the coincidence limit [cf. Eq.~(\ref{eq:CA})], 
the Fourier component is defined as 
\begin{eqnarray}
\tilde W(p;s) \equiv \int d^4x \ \e^{ipx} W(x;s)  
\, ,
\end{eqnarray}
where $ x^\mu = x_1^\mu-x^\mu_0 $. 
Below, we closely look into the structure of the amplitude $ \tilde W(p;s)  $ in the momentum-space representation.

The ``Schr\"odinger equation'' in the momentum space reads 
\begin{eqnarray}
\label{eq:dK-p}
i \frac{ d \tilde W(p;s) }{ds} = \left[ \, - \partial_p K \partial_p  + \frac{1}{ 2 i}  \tr[ (q_f F)  \coth (q_f F s)]  \, \right] \tilde W(p;s)
\, ,
\end{eqnarray}
with $  K$ defined below Eq.~(\ref{eq:Ham-bilinear}). 
Notice that the derivative operator on the right-hand side is quadratic. 
Therefore, we may put an Ansatz \cite{Brown:1975bc, Duff:1975ue, Dittrich:1985yb} 
\begin{eqnarray}
\label{eq:ansatz}
\tilde W(p;s) =  C_p \exp\left( i p X p + Y \right)
\, ,
\end{eqnarray}
where the symmetric tensor $ X^{\mu\nu}(s) $ and the scalar function $Y(s)  $ will be determined below. 
Inserting the Ansatz into Eq.~(\ref{eq:dK-p}), we have 
\begin{eqnarray}
\left[ \, p \left( 4 X K X + \frac{ dX}{ds} \right)p 
- i \left( 2 \tr[KX] + \frac{1}{2} \tr[(q_f F)  \coth (q_f F s)] +   \frac{ dY}{ds}  \right)\, \right] \tilde W(p;s) = 0
\, .
\end{eqnarray}
Therefore, we get a system of coupled equations 
\begin{subequations}
\begin{eqnarray}
&&
 4 X K X + \frac{ dX}{ds} = 0
 \, ,
 \\
 &&
2  \tr[KX]  + \frac{1}{2} \tr[(q_f F)  \coth (q_f F s)]  +  \frac{ dY}{ds} = 0
\, .
\end{eqnarray}
\end{subequations}
With the help of the basic properties of the hyperbolic functions, 
we can convince ourselves that the following functions satisfy those equations: 
\begin{subequations}
\begin{eqnarray}
&&
X^{\mu\nu} = [ (q_f F)^{-1} \tanh(q_fFs) ]^{\mu\nu}
\, ,
\\
&&
Y(s) = - \frac{1}{ 2} \tr [ \ln\{ \cosh(q_f Fs) \} ]
\, .
\end{eqnarray}
\end{subequations}

As we have done in the previous subsection, we can diagonalize $  F_\mu^{\  \nu}  $ to find 
\begin{eqnarray}
\e^Y = {\rm det} [ \cosh(q_f Fs)  ]^{-\frac{1}{2}} 
= \frac{1}{  \cosh (q_f a s) \cos (q_f b s)  }
\, .
\end{eqnarray}
Likewise, $ X_\mu^{\ \, \nu} $ can be diagonalized by a unitary matrix $ M_\mu^{\ \, \nu}  $. 
We shall choose the unitary matrix so that we have $ M^{-1} F M = \diag ( a, ib, -i b, -a )$. 
Then, the bilinear form can be diagonalized as 
\begin{eqnarray}
p X p = \frac{  \tanh(q_fas)}{q_f a} p^{\prime \, 2}_\para  +  \frac{   \tan(q_f b s)}{q_f b} p^{\prime \, 2}_\perp
\, .
\end{eqnarray}
On the right-hand side, we defined the transformed momentum $  p^\prime_\mu = (M^{-1})_\mu^{\ \, \nu} p^\nu$, 
and further $  p_\para^{ \mu} := (p^0,0,0,p^3)$ and $ p^\mu_\perp := (0,p^1,p^2,0) $ in this basis (written without the primes).\footnote{
One may wonder the meaning of the symbols $ \para, \perp $. 
They actually refer to the directions parallel and perpendicular to the magnetic field 
in such a configuration that $ a=0 $ and $ b=B $. 
We just follow these notations familiar in a certain community, 
although we here do not need to specify a specific configuration of the external field or a specific Lorentz frame. 
}

With the $ X^{\mu\nu} $ and $Y  $ determined above, 
we should have the normalization $ C_p =1 $ to reproduce the free result (\ref{eq:W0}) in the vanishing field limit $ a,b\to0 $. 
Therefore, noting that $  \det M =1$ and dropping the prime on the momentum, 
we find the momentum-space representation of the HE Lagrangian 
\begin{subequations}
\label{effectiveaction-p}
\begin{eqnarray}
{\cal L}_s^{(1')} &=& 
- i \int_0^\infty \frac{ds}{s}\,  {\rm e}^{-is(m^2-i\epsilon)} 
\pint
\left[
\frac{ \e^{  i \frac{  \tanh(q_fas)}{q_f a} p_\para^2  + i \frac{   \tan(q_f b s)}{q_f b} p_\perp^2  }   } 
{ \cosh (q_f a s) \cos (q_f b s) } 
- \e^{ i p^2 s}
\right] 
\, ,
\\
{\cal L}_f^{(1')} &=& 
2i \int_0^\infty \frac{ds}{s}\,  {\rm e}^{-is(m^2-i\epsilon)} 
\pint
\left[ 
\e^{  i \frac{  \tanh(q_fas)}{q_f a} p_\para^2  + i \frac{   \tan(q_f b s)}{q_f b} p_\perp^2  } 
 - \e^{ i p^2 s}
\right] 
\, .
\end{eqnarray}
\end{subequations}

\subsection{Decomposition into the Landau levels}

With the momentum-space representation (\ref{effectiveaction-p}), 
we are in position to investigate the Landau-level structures of the HE effective Lagrangians. 
In both scalar and spinor QED, we deal with the similar momentum integrals 
\begin{subequations}
\begin{eqnarray}
&&
I_p^s(a,b) \equiv \int \frac{d^4p}{(2\pi)^4} \frac{1}{ \cosh (q_f a s) \cos (q_f b s) } \exp\bigg[
 i\frac{p_\parallel^2}{q_fa}\tanh(q_fa s) + i\frac{p_\perp^2}{q_fb}\tan(q_fbs)
\bigg]
,
\\
&&
I_p^f(a,b) \equiv 2 \int \frac{d^4p}{(2\pi)^4} \exp\bigg[
 i\frac{p_\parallel^2}{q_fa}\tanh(q_fa s) + i\frac{p_\perp^2}{q_fb}\tan(q_fbs)
\bigg]
,
\end{eqnarray}
\end{subequations}
where the first and second ones are for scalar and spinor QED, respectively. 
Notice that the four-dimensional integral is completely factorized into 
the longitudinal ($ p_\para $) and transverse ($ p_\perp $) momentum integrals: 
Each of them is a two-dimensional integral. 
We first perform the longitudinal-momentum ($ p_\para $) integrals. 
The Gaussian integrals are straightforwardly carried out as 
\begin{subequations}
\begin{eqnarray}
I_p^s(a,b) 
&=& I_\perp^s (b)
\left[  \frac{  i}{(2\pi)^2} \frac{\pi q_f a }{ i \sinh(q_f as)} \right]
\label{eq:K-scalar-Landau}
\, ,
\\
I_p^f(a,b) 
&=&I_\perp^f (b)
\left[  \frac{  i}{(2\pi)^2} \frac{\pi q_f a }{ i \tanh(q_f as)} \right]
\label{eq:K-fermion-Landau}
\, .
\end{eqnarray}
\end{subequations}
An imaginary unit $i$ arises from the Wick rotation of the temporal component. 
Here, we have written the remaining transverse-momentum part $ (p_\perp) $ as 
\begin{subequations}
\label{eq:I_perp}
\begin{eqnarray}
I_\perp^s (b)&\equiv& \int \frac{d^2p_\perp}{(2\pi)^2} \frac{ 1 } { \cos (q_f b s) } 
 \exp\bigg[i\frac{p_\perp^2}{q_fb}\tan(q_fbs)\bigg]
\, ,
\\
I_\perp^f (b) &\equiv& 2 \int \frac{d^2p_\perp}{(2\pi)^2}  
 \exp\bigg[i\frac{p_\perp^2}{q_fb}\tan(q_fbs)\bigg]
\, .
\end{eqnarray}
\end{subequations}
Performing the remaining Gaussian integral as well, 
one can straightforwardly reproduce the previous result (\ref{effectiveaction-x}).

Here, before performing the transverse-momentum integral, 
we carry out the Landau-level decomposition 
by the use of the generating function of the associated Laguerre polynomial 
\begin{eqnarray}
(1-z)^{-(1+\alpha)} \exp \left( \frac{xz}{z-1} \right) 
= \sum_{n=0}^{\infty} L_n^\alpha (x) z^n
\label{eq:g_Laguerre}
\, .
\end{eqnarray}
To do so, we put 
\begin{eqnarray}
z = - \e^{-2i \vert q_f b \vert s}
\label{eq:zzz}
\,  .
\end{eqnarray}
Then, the tangent in the exponential shoulder is rewritten in a desired form 
\begin{eqnarray}
\label{eq:1}
\exp\Big( \, i \frac{ p_\perp^2 }{ q_f b } \tan(  q_f b  s) \, \Big)
= \exp\Big(\, - \frac{ u_\perp }{ 2 } \, \Big) \exp\Big(\, \frac{ u_\perp z}{z-1} \, \Big)
\, ,
\end{eqnarray}
with $ u_{\perp}= -2 p_\perp^2 /\vert q_f b \vert$. 
Also, the other trigonometric function is also arranged as $ \cos(q_f b s) = ( 1 - z) (-z)^{-1/2}/2 $. 
Identifying the exponential factors in Eqs.~(\ref{eq:g_Laguerre}) and (\ref{eq:1}), 
the $ p_\perp $-dependent part is decomposed as 
\begin{subequations}
\begin{eqnarray}
I_\perp^s (b)&=& 2 \sum_{n=0}^{\infty} \e^{ -i |q_f b| (2n+1) s}   (-1)^n
\int \frac{d^2p_\perp}{(2\pi)^2} L_n (u_\perp) \e^{- \frac{ u_\perp }{ 2 } }
\, ,
\\
I_\perp^f (b) &=& 2 \sum_{n=0}^{\infty} \e^{ -2i |q_f b| n s}  (-1)^n
\int \frac{d^2p_\perp}{(2\pi)^2} L_n^{-1} (u_\perp) \e^{- \frac{ u_\perp }{ 2 }}
\, .
\end{eqnarray}
\end{subequations}
The additional factor of 2 in the scalar case, as compared to Eq.~(\ref{eq:I_perp}), 
comes from the expansion of cosine factor.
Performing the transverse-momentum integrals as elaborated in Appendix~\ref{sec:I_perp}, 
we obtain quite simple analytic results 
\begin{subequations}
\begin{eqnarray}
I_\perp^s (b) &=& 
\sum_{n=0}^\infty  \left[ \frac{|q_f b|}{2\pi} \right]  \e^{ - i |q_f b| (2n+1) s}
\, ,
\\
I_\perp^f (b) &=& 
\sum_{n=0}^\infty  \left[\kappa_\lambda \frac{|q_f b|}{2\pi} \right]  \e^{ -2i |q_f b| n s}
\, ,
\end{eqnarray}
\end{subequations}
where  $ \kappa_n = 2 - \delta_{n0} $. 
The results of the transverse-momentum integrals are 
independent of the index $ n $ up to the dependence in $ \kappa_n $.

\begin{figure}[t]
     \begin{center}
              \includegraphics[width=0.7\hsize]{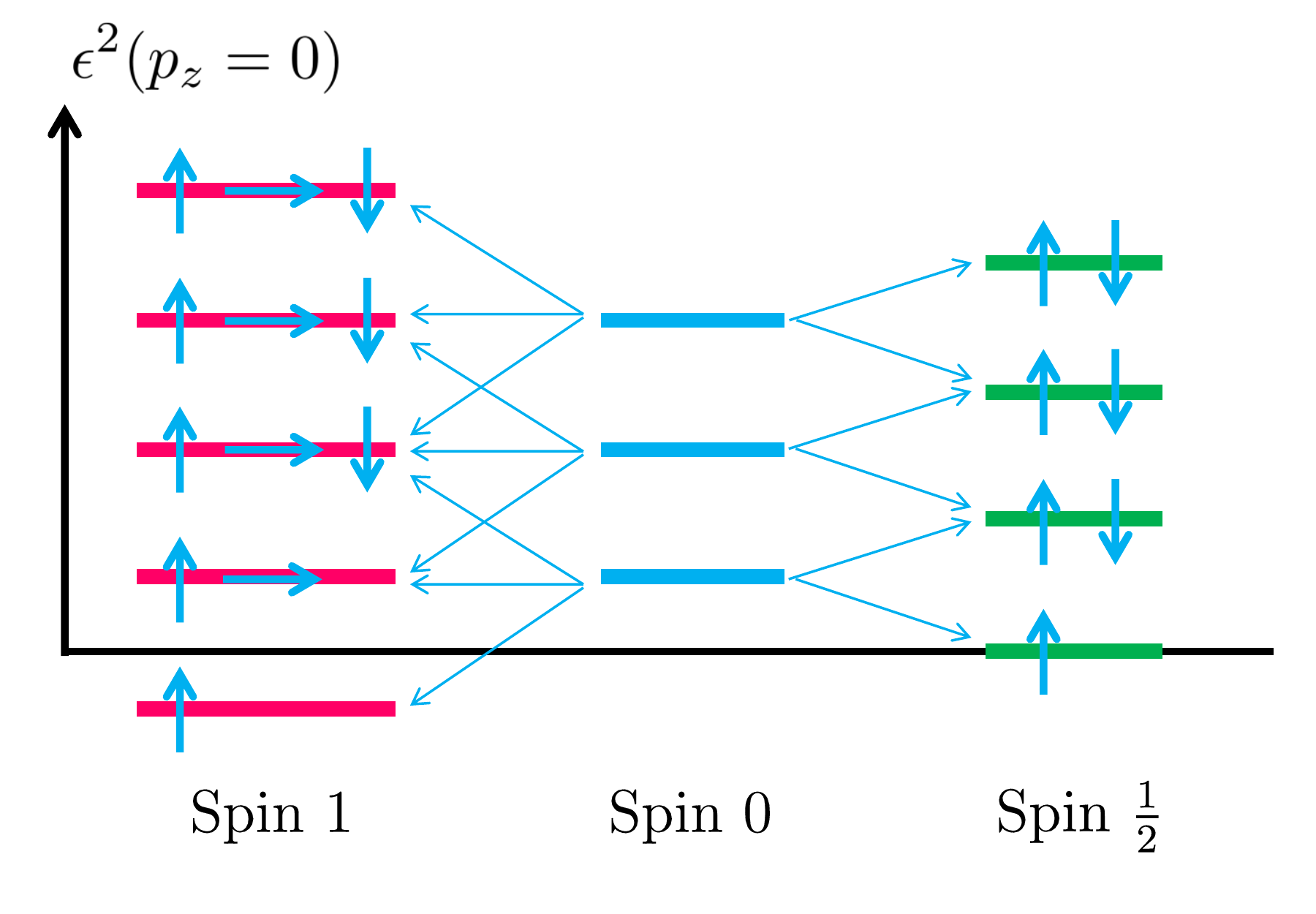}
     \end{center}
\vspace{-1cm}
\caption{
The resultant energy spectra from the relativistic Landau quantization and the Zeeman splitting with the $ g $-factor, $ g=2 $. 
While the ground state energy of the spinless particles is given by the ``zero-point energy'' of the Landau quantization, 
those for the spinning particles are shifted by the Zeeman effect. 
}
\label{fig:spectrum}
\end{figure}

Plugging the above integrals into Eq.~(\ref{effectiveaction-p}), the effective Lagrangian is obtained as 
\begin{subequations}
\label{eq:HE-Landau}
\begin{eqnarray}
&&
{\cal L}_{\rm HE}^s =
\sum_{n=0}^\infty \left[  \frac{|q_f b|} {2\pi} \right]
\left[ - \frac{i}{4\pi}\int_0^\infty \frac{ds}{s} \e^{-i \{ (m_n^{s})^{2} - i \epsilon\} s}   \frac{ q_f a }{ \sinh(q_f a s) } \right]
\, , 
\label{HE-Landau-s}
\\
&&
{\cal L}_{\rm HE}^f =
 \sum_{n=0}^\infty \left[ \kappa_n \frac{|q_f b|} {2\pi} \right]
\left[ \frac{i}{4\pi}\int_0^\infty \frac{ds}{s} \e^{-i \{ (m_n^{f})^{2} - i \epsilon\} s}   \frac{ q_f a }{ \tanh(q_f a s) } \right]
\, .
\label{HE-Landau-f}
\end{eqnarray}
\end{subequations} 
We have defined the effective masses 
\begin{subequations}
\label{eq:mass-LL}
\begin{eqnarray}
&&
( m_n^{s})^{2} = m^2 + (2 n+1) | q_f b|  
 \,  ,
 \\
 &&
(  m_n^{f})^{2} = m^2 + 2 n |q_f b|
  \, .
\end{eqnarray}
\end{subequations}
Note that we could drop the $ \epsilon $ parameter in the scalar QED result (\ref{HE-Landau-s}), 
because the integrand at each $  n$ is regular along the integral contour 
and is convergent asymptotically.

Remarkably, the one-loop correction to the effective action 
appears in the relativistic form of the Landau levels specified by the integer index $ n $. 
As long as $  b \not = 0$, there exists such a Lorentz frame that 
this Lorentz invariant reduces to the magnetic-field strength $ b = |\bB| $. 
Accordingly, we can identify the Landau levels in such a frame. 
The reason for the difference between the boson and fermion spectra 
is the additional Zeeman shift which depends on the spin size (cf. Fig.~\ref{fig:spectrum}).  
This interpretation is justified by tracking back the origin of the difference. 
Remember that scalar QED has the cosine factor in Eq.~(\ref{eq:I_perp}) 
which results in the factor of $ z^{1/2} $ [cf. expansion below Eq.~(\ref{eq:1})] 
and then the ``zero-point energy'' of the Landau level. 
In spinor QED, this cosine factor is cancelled by the spin-interaction term (\ref{eq:F-sigma}).

In both scalar and spinor QED, the results are given as the sum of the independent contribution from each Landau level. 
Moreover, the two-fold degenerated spin states in the higher levels, 
seen in Fig.~\ref{fig:spectrum}, provide the same contributions. 
Those properties would be specific to the one-loop results, 
and may be changed in the higher-loop contributions where 
the dynamical photons could induce the inter-level transitions and also ``probe'' the spin states. 
Importantly, the transverse-momentum integrals yield the Landau degeneracy factor $|q_f b|/2\pi  $ 
between the first square brackets in Eq.~(\ref{eq:HE-Landau}). 
Since all the Landau levels have the same degeneracy, 
it is anticipated that this factor is independent of $ n $.  
An additional spin degeneracy factor $ \kappa_n $ automatically appears in spinor QED 
as a result of the transverse-momentum integral (cf. Appendix~\ref{sec:I_perp}).

Now, one can confirm that the proper-time integrals between the second square brackets in Eq.~(\ref{eq:HE-Landau}) 
exactly agree with the HE Lagrangian in the (1+1) dimensions 
for the particles labelled with the effective mass $ m_n $. 
Note that the powers of $ 1/\pi $, $  i$, $  s$ are different from 
those in the familiar four-dimensional effective actions (\ref{effectiveaction-x}), 
because those factors depend on the spatial dimensions and the proper time is a dimensionful variable.

Note also that one may not consider the vanishing limit $ b \to 0$ before 
taking the summation over $ n $ in Eq.~(\ref{eq:HE-Landau}), 
since the summation and the limit do not commute with each other. 
A finite Landau spacing should be maintained in the summation form 
so that the spectrum tower does not collapse into the ground state. 
The convergence of the summation should faster for a larger $ |b| $ 
where the Landau spacing becomes large. 

Summarizing, we have found that the HE effective Lagrangian 
can be decomposed into the simple summation form with respect to the Landau levels. 
In fact, one can directly obtain the same results from the well-known forms of 
the HE effective action (\ref{effectiveaction-x}) by the use of identities:
\begin{subequations}
\label{eq:trigo}
\begin{eqnarray}
&&
\sec x 
= 2i \frac{  \e^{-i  x} }{1-\e^{-2i  x} }  
= 2i \sum_{n=0}^\infty \e^{-i (2n +1) x}  
\, ,
\\
&&
\cot x 
= i\left[1+ \frac{ 2 \e^{-2i  x} }{1-\e^{-2i  x} }  \right]
= i \sum_{n=0}^\infty \kappa_n\, \e^{-2i n x}  
\, .
\end{eqnarray}
\end{subequations}
At the one-loop order, there is no mixing among the Landau levels 
and the HE effective action is given by the sum of independent Landau-level contributions. 
In each Landau-level contribution, the effective Lagrangian is given as the product of 
the Landau degeneracy factor and the HE effective Lagrangian in the (1+1)-dimensions.

\subsection{Schwinger mechanism in the Landau levels}

\begin{figure}
     \begin{center}
              \includegraphics[width=0.6\hsize]{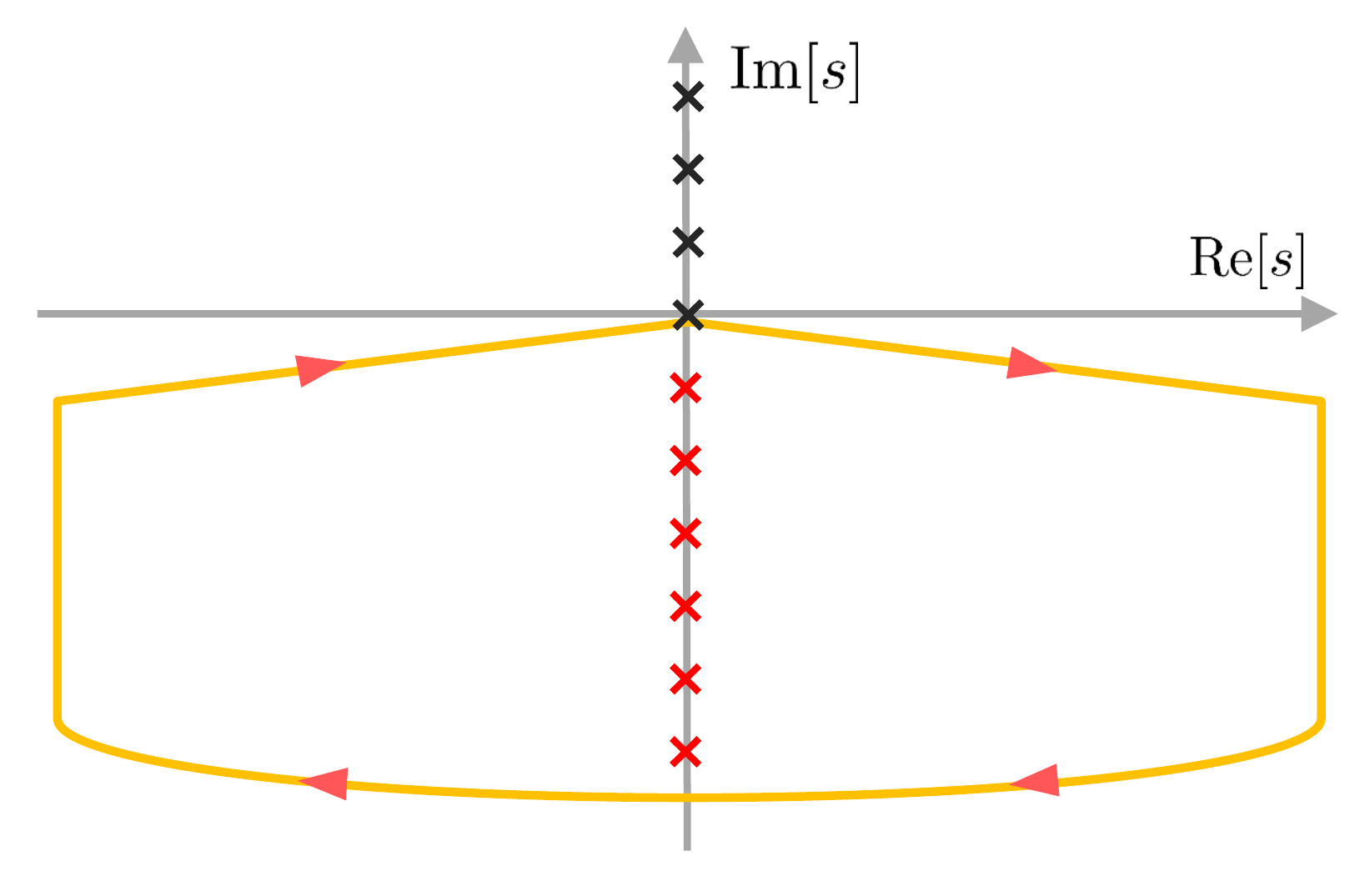}
     \end{center}
\vspace{-1cm}
\caption{
Pole structures in the proper-time representation. 
The Landau-level representations (\ref{eq:HE-Landau}) have poles only on the imaginary axis. 
}
\label{fig:contour}
\end{figure}

It has been known that the HE effective action acquires an imaginary part in an electric field, 
implying creation of on-shell particles out of, otherwise, virtual states forming bubble diagrams in vacuum. 
While the real part of the HE effective action describes electromagnetism, 
the production of particle and antiparticle pairs in the electric fields is 
signalled by the emergence of an imaginary part \cite{Sauter:1931zz, Heisenberg:1935qt}. 
This is often called the Schwinger mechanism \cite{Schwinger:1951nm}. 
Here, we compute the imaginary part on the basis of the Landau-level representation (\ref{eq:HE-Landau}).\footnote{
The imaginary part of the effective action provides the vacuum persistence probability, 
which is, by definition, different from the pair production rate. 
The latter should be computed as the expectation value of 
the number operator (see, e.g., Refs.~\cite{Nikishov:1969tt, Holstein:1999ta, Cohen:2008wz, Tanji:2008ku, Gelis:2015kya}). 
}

Since the integrands in the effective actions (\ref{eq:HE-Landau}) are even functions of $ s $, 
the imaginary part of the proper-time integral can be written as 
\begin{eqnarray}
\Im m \left[ i \int_0^\infty ds \, {\rm e}^{-i(m^2-i\epsilon)s} f(s) \right]
&=& \frac{1}{2} \left[ \, \int _0^\infty  ds \, {\rm e}^{-i(m^2-i\epsilon)s} f(s) -  \int_0^\infty ds \, {\rm e}^{i(m^2 + i\epsilon)s} f(s)  \right]
\nn
\\
&=& \frac{1}{2} \int_{-\infty}^\infty ds \, {\rm e}^{-i(m^2-i \sgn(s) \epsilon)s} f(s) 
\, ,
\end{eqnarray}
where $ f(s) $ is the even real function. 
Based on the last expression, one can consider the closed contour: 
Because of the infinitesimal parameter $ \epsilon $, the contour along the real axis is inclined below the axis, 
which, together with the positivity of $ m_\lambda ^2$, suggests rotating the contour into the lower half plane (cf. Fig.~\ref{fig:contour}).


One should notice the pole structures arising from the hyperbolic functions in the effective actions (\ref{eq:HE-Landau}). 
When $ a \not = 0 $, there are an infinite number of poles on the imaginary axis 
located at $  s =  i n\pi/|q_f a| =:  i s_n $ with $ n \in Z$. 
Therefore, picking up the residues of those poles, 
we obtain the imaginary parts of the HE effective actions 
\begin{subequations}
\label{imaginary_E-Landau}
\begin{eqnarray}
&&
\Im m {\cal L}_{\rm HE}^s = 
\sum_{n=0}^\infty \left[  \frac{|q_f b|} {2\pi} \right]
 \sum_{\sigma=1}^\infty (-1)^{\sigma-1} \frac{ |q_f a| }{4\pi \sigma}  \e^{- m_n^{s2} s_\sigma} 
\, , 
 \label{imaginary_E-Landau-s}
\\
&&
\Im m {\cal L}_{\rm HE}^f =
 \sum_{n=0}^\infty \left[ \kappa_n \frac{|q_f b|} {2\pi} \right]
 \sum_{\sigma=1}^\infty \frac{ |q_f a| }{4\pi \sigma}  \e^{- m_n^{f2} s_\sigma} 
\, .
 \label{imaginary_E-Landau-f}
\end{eqnarray}
\end{subequations}
Those imaginary parts indicate the vacuum instability in the configurations with finite values of $ a $. 
This occurs only in the presence of an electric field and is interpreted as an instability due to a pair creation from the vacuum 
as known as the Schwinger mechanism~\cite{Schwinger:1951nm}. 
Note that, after the subtraction of the free-theory contribution in Eq.~(\ref{effectiveaction-x}), 
there is no contribution from the pole on the origin, meaning that this pole is nothing to do with the Schwinger mechanism. 
In the Landau-level representation (\ref{eq:HE-Landau}), the subtraction of the free-theory contribution is somewhat tricky 
because one cannot take the vanishing $ b $ limit before taking the summation as mentioned above Eq.~(\ref{eq:trigo}). 
Nevertheless, the second-rank pole at the origin does not contribute to the integral.

Now, we fix the magnitude of the electric field $ |\bE| $, and investigate 
how a magnetic field modifies the magnitude of the imaginary part as compared to the one 
in a purely electric field. To see a dependence on the relative direction 
between the electric and magnetic fields, we consider two particular 
configurations in which those fields are applied in parallel/antiparallel and orthogonal to each other. 
The covariant expression for the general field configuration provides the interpolation between those limits.

When a magnetic field is applied in {\it parallel/antiparallel} to 
the electric field, we have $a=|\bE|$ and $b=|\bB|$. Compared with 
the purely electric field configuration, we get a finite $b$ without changing the value of $ a $. 
We may thus define the critical electric field, where the exponential factor reduces to order one ($ m_n^2 s_1 = 1 $), 
with the energy gap of each Landau level as $ E_n^c \equiv m_n^2/|q_f| $. 
Therefore, there is an infinite number of the ``Landau-Schwinger limits'' specifying 
the critical field strengths for the pair production at the Landau-quantized spectrum. 
It is quite natural that the exponential suppression is stronger for 
the higher Landau level which has a larger energy gap measured from the Dirac sea. 
However, once we overcome the exponential suppression with a sufficiently strong electric field, 
the magnitude of the imaginary part turns to be enhanced by the Landau degeneracy factor. 
This is because an energy provided by the external electric field can be consumed only to fill up 
the one-dimensional phase space along the magnetic field, 
and the degenerated transverse phase space can be filled without an additional energy cost.

Remarkably, the lowest critical field strength for the {\it fermions}, $ E^c_{n=0} = m^2/|q_f| $, 
is independent of the magnetic field. 
Therefore, the parallel magnetic field {\it catalyzes} the Schwinger pair production thanks to the LLL contribution.  
The origin of this enhancement is the aforementioned effective dimensional reduction, 
and is somewhat similar to the ``magnetic catalysis'' of the chiral symmetry breaking \cite{Gusynin:1994xp, Gusynin:1995nb}.\footnote{Note, however, that the magnetic catalysis 
is more intimately related to the effective low dimensionality of the system  
rather than just the enhancement by the Landau degeneracy factor 
(see, e.g., Refs.~\cite{Fukushima:2012xw, Hattori:2017qio}).
}  
However, the lowest critical field strength for the {\it scalar} particles increases 
as we increase the magnetic field strength according to the spectrum (\ref{eq:mass-LL}). 
Therefore, the Schwinger mechanism is suppressed with the parallel magnetic field. 
Besides, the scalar QED result (\ref{imaginary_E-Landau-s}) is given as the alternate series. 
The spinor QED result (\ref{imaginary_E-Landau-f}) does not have the alternating signs 
because of the additional hyperbolic cosine factor from the spin-interaction term in the numerator. 
Therefore, the alternating signs originate from the quantum statistics. 
We will find that the gluon contribution, as an example of vector bosons, 
is also given as an alternate series in the next section.

When a magnetic field is applied in {\it orthogonal} to the electric field, i.e., when $\G=0$ (with $ \F \not = 0 $), 
any field configuration reduces to either a purely electric or magnetic field by a Lorentz transform. 
When $\F\geq0$, i.e., $|\bE| \leq |\bB|$, we have $a=0$. 
Therefore, the imaginary parts vanishes and no pair production occurs. 
When $\F<0$, i.e., $|\bB| < |\bE|$, we have $a = \sqrt{|\bE^2 - \bB^2|} $ and $ b=0 $. 
In this case, the summation formula is not useful as discussed above Eq.~(\ref{eq:trigo}). 
Instead, we should rely on the original Schwinger's formula \cite{Schwinger:1951nm}, 
where we observe a pair production induced by a purely electric field with a strength $  a$. 
Because of a smaller value  $a < |\bE|$, the imaginary part is suppressed by the magnetic field. 
In the presence of an orthogonal magnetic field, a fermion and antifermion drift in the same direction 
perpendicular to both the electric and magnetic fields. 
This cyclic motion prevents the pair from receding from each other along the electric field, 
which may cause a suppression of the pair production.


\section{QCD in covariantly constant chromo fields}

\label{sec:QCD} 

In this section, we extend the HE effective action to 
its counterpart for QCD in external chromo-electromagnetic fields. 
We first briefly capture the QCD Lagrangian in the external chromo-electromagnetic field
on the basis of the ``background-field method.'' 
We shall start with the full QCD action with the SU$(N_{c})$ color symmetry: 
\beq
S_{\rm QCD}
&=& \int d^{4}x \Big[
\bar{\psi} \left( i \sla D_{\mathcal{A}} - m \right) \psi 
-\frac{1}{4} \mathcal{F}_{\nonA \mu \nu}^{a} \mathcal{F}^{a \mu \nu}_\nonA  
\Big]
\, .
\label{QCDaction}
\eeq
The covariant derivative here is defined with the non-Abelian guage field as 
\beq
D^{\mu}_{\mathcal{A}} = \partial^{\mu} - ig\mathcal{A}^{a \mu} t^{a} 
\, .
\label{QCDcovderivative}
\eeq
The associated field strength tensor is given by
$
\mathcal{F}^{a \mu \nu}_\nonA 
= \partial^{\mu} \mathcal{A}^{a \nu} - \partial^{\nu} \mathcal{A}^{a \mu} + gf^{abc} \mathcal{A}^{b \mu} \mathcal{A}^{c \nu} $. 
The generator of the non-Ableian gauge symmetry obeys the algebra 
$[ t^a, t^b] = i f^{abc} t^c$ and $\tr( t^a t^b ) = C \delta^{ab}$ 
with $C =1/2  $ and $C= N_c=3  $ for the fundamental and adjoint representations, respectively. 
While we consider one-flavor case for notational simplicity, 
extension to multi-flavor cases is straightforward.

We shall divide the non-Ableian gauge field into the dynamical and external fields: 
\begin{eqnarray}
\nonA^{a\mu} = a^{a\mu} + \nonA_\ext^{a\mu}
\,  .
\label{eq:aA}
\end{eqnarray} 
Then, the kinetic terms read 
\begin{eqnarray}
\Lag_{\rm kin}  &=& 
\bar \psi (i \slashed D-m) \psi  - \bar c^a (D^2)^{ac} c^c 
\nn
\\
&& 
- \frac{1}{2} a_\mu^a \left( 
- (D^2)^{ac} g^{\mu\nu}  + ( 1-  \frac{1}{\xi_{g} } )  D^{ ab \mu} D^{ bc\nu} 
+ i g ( \nonF_{\alpha\beta}^b \J^{\alpha\beta})^{\mu\nu} f^{abc} 
\right) a_\nu^c
 \, ,
 \label{eq:Lkg_d}
\end{eqnarray}
where the covariant derivative is 
defined with the external chromo-field $  D^{ \mu} \equiv  \partial^\mu - ig \nonA_\ext^{a\mu} t^a$. 
The ghost field and the gauge parameter (for the dynamical gauge field) are denoted as $ c $ and $ \xi_g $, respectively. 
We also introduced the field strength tensor of the external field 
$ \nonF^{a \mu\nu}\equiv \partial^\mu \nonA^{a \nu }_\ext - \partial^\nu \nonA^{a\mu}_\ext
 - i g (t^b)^{ac} \nonA^{b\mu}_\ext \nonA^{c\nu}_\ext$ 
and the generator of the Lorentz transformation $\J_{\alpha\beta}^{\mu\nu} 
 = i ( \delta_\alpha ^\mu \delta_\beta^\nu - \delta_\beta^\mu \delta_\alpha^\nu )$ 
 so that $( \nonF_{\alpha\beta}^b \J^{\alpha\beta})^{\mu\nu}= \nonF^{b\alpha\beta} \J_{\alpha\beta}^{\mu\nu}=2i\nonF^{b\mu\nu}$.

\subsection{Covariantly constant chromo fields}

While we have not assumed any specific configuration of the external fields in the above arrangement, 
we now focus on the covariantly constant external field. 
It is an extension of the constant Abelian electromagnetic field that satisfies the covariant condition 
\cite{Batalin:1976uv, Yildiz:1979vv, Ambjorn:1982bp, 
Gyulassy:1986jq, Suganuma:1991ha, Nayak:2005yv, Nayak:2005pf, Tanji:2010eu, Ozaki:2013sfa, Ozaki:2015yja} 
\begin{eqnarray}
D^{ab}_\lambda \nonF^b_{\mu\nu} = 0
\label{eq:CCF}
\,  .
\end{eqnarray}
As shown in Appendix~\ref{sec:covariantly-constant}, 
we find the solution in a factorized form 
\begin{eqnarray}
\nonF^a_{\mu\nu} = \nonF_{\mu\nu} n^a
\label{eq:fact}
\, ,
\end{eqnarray}
where $n^a$ is a vector in the color space 
and is normalized as $n^a n^a = 1$. 
The vector $ n^a $ represents the color direction, 
while an Abelian-like field $\nonF^{\mu\nu}$ quantifies the magnitude of the external field. 
We define the Poincar\'e invariants $ \aaa $ and $\bbb  $ as in Eq.~(\ref{eq:Poincare}) 
with the field strength tensor $\nonF^{\mu\nu}$.

According to the above factorization, the external gauge field in the covariant derivative is also 
factorized into the color direction and the magnitude. 
Then, the color structures in the covariant derivatives are diagonalized as (cf. Appendix~\ref{sec:covariantly-constant})
\begin{subequations}
\begin{eqnarray}
D^{ij \mu} &=&  \delta^{ij} \left(  \partial^\mu - i w_i  { \nonA}_\ext ^\mu \right)
\label{eq:fund_ccf}
\, ,
\\
D^{ab \mu} &=&  \delta^{ab}  \left(  \partial^\mu - i v^{a} \nonA^\mu_\ext  \right) 
\label{eq:adj_ccf}
\, ,
\end{eqnarray}
\end{subequations}
where the summation is not taken on the right-hand sides 
and the first and second lines are for the fundamental and adjoint representations, respectively. 
The effective coupling constants $ w_i $ and $v^a  $ are specified 
by the second Casimir invariant of the color group \cite{Batalin:1976uv, Yildiz:1979vv, Ambjorn:1982bp, 
Gyulassy:1986jq, Suganuma:1991ha, Nayak:2005yv, Nayak:2005pf, Tanji:2010eu, Ozaki:2013sfa, Ozaki:2015yja}. 
In the same way, we also get the diagonal form of the spin-interaction term 
\begin{eqnarray}
ig (\nonF_{\alpha\beta}^b \J^{\alpha\beta})^{\mu\nu} f^{abc}  
= 
v^a \delta^{ac}(\nonF_{\alpha\beta} \J^{\alpha\beta})^{\mu\nu} 
\label{eq:spin-gluon}
\, .
\end{eqnarray}
Below, we investigate the HE effective Lagrangian in the covariantly constant chromo fields.

\subsection{Effective actions in the covariantly constant chromo fields}

Since the color structures are diagonalized in the covariantly constant fields, 
the effective Lagrangians are composed of the sum of the color indices 
\begin{subequations}
\label{eq:HE-YM}
\begin{eqnarray}
\Lag_\quark &=& \sum_{i=1}^3 \left.  \Lag_f \right|_{q_f \to - w_i}
\, ,
\\
\Lag_\ghost &=& -  \sum_{a=1}^8  \left.   \Lag_s \right|_{q_f \to - v^a}
\, ,
\\
\Lag_\gluon &=& \sum_{a=1}^8 \Lag_g^{a} 
\, .
\end{eqnarray}
\end{subequations}
The contributions from the fermion loop $ \Lag_f $ and scalar loop $\Lag_s$ 
have been computed in the previous section at the one-loop order. 
The quark and ghost contributions can be simply obtained by replacing the charges 
and attaching a negative overall sign to the scalar QED contribution 
for the Grassmann nature of the ghost field. 
Therefore, we only need to compute the gluon contribution below. 
Moreover, since the contributions with eight different colors are just additive to each other, 
we may focus on a particular color charge. 
For notational simplicity, we drop the color index on $  v^a$ and $ \Lag_g^{a}  $ below, 
and write them $  v$ and $ \Lag_g^{(1)}  $ for the one-loop order, respectively.

Performing the path integration over the gluon bilinear field, 
we again start with the determinant 
\beq
S^{(1)}_g [A_\mu]= \frac{i}{2} \ln \det [ D^2 g^{\mu\nu} - v (\nonF_{\alpha\beta} \J^{\alpha\beta})^{\mu\nu} ] 
\label{Lag_original-g}
\, .
\eeq 
With the aid of $  \det(-1)=1$ in the four-dimensional Lorentz index, we let the sign in front of $  D^2$ positive. 
Note that we have dropped all the diagonal color indices as promised above. 
The proper-time representation is immediately obtained as [cf. Eq.~(\ref{propertime2})] 
\beq
{\cal L}_g^{(1')}= - \frac{i}{2}\int_0^\infty \frac{ds}{s}\,  {\rm e}^{-\epsilon s} \, 
{\rm tr}\, \left[ 
\langle x | {\rm e}^{-i \hat H_g^{\mu\nu}  s}|x\rangle - \langle x|{\rm e}^{-i \hat H_{g0}^{\mu\nu}  s}|x\rangle 
\right] 
\, ,
\label{effectiveaction-g}
\eeq
where  ``tr" indicates the trace over the Lorentz indices. 
The ``Hamiltonian" for the gluon contribution may be defined as 
\beq
\hat H_g^{\mu\nu} \equiv  D^2 g^{\mu\nu} - v^a(\nonF_{\alpha\beta} \J^{\alpha\beta})^{\mu\nu} 
\label{Ham_gluon}
\, ,
\eeq 
and $ \hat H_{g0} ^{\mu\nu}  =  g^{\mu\nu} \del^2 \ (=   g^{\mu\nu} \hat H_{s0})$.


Compared with the fermionic Hamiltonian (\ref{Ham_fermion}), 
the spin-interaction term is replaced by $  (\nonF_{\alpha\beta} \J^{\alpha\beta})^{\mu\nu} $, 
that is, the field strength tensors are coupled with the Lorentz generators 
in the spinor and vector representations, respectively. 
After the Landau-level decomposition, we will explicitly see 
the dependence of the Zeeman effect on the Lorentz representations. 
Recall that the (Abelian) field strength tensor $ \nonF_\mu^{\ \, \nu} $ can be diagonalized 
as $ (M^{-1} \nonF M)_\mu^{\ \, \nu} = \diag( \aaa, i\bbb, -i\bbb, -\aaa) $. 
Therefore, one can easily carry out the trace 
\begin{eqnarray}
\label{eq:spin-g}
{\rm tr}\, \left[ \langle x | {\rm e}^{-i \hat H_g^{\mu\nu}  s}|x\rangle \right]
&=& \tr \left[  \e^{- 2 s v \nonF_\mu^{\ \, \nu} } \right]
 \langle x | {\rm e}^{+ i \hat H_s  s}|x\rangle 
\nn
\\
&=& 2 [ \cosh(2 v \aaa s) +  \cos(2 v \bbb s) ]
 \langle x | {\rm e}^{- i \hat H_s  s}|x\rangle 
\, .
\end{eqnarray}
Then, we are again left with the transition amplitude which has the same form as in scalar QED 
(with appropriate replacements of the field strength tensor and the charge). 
Plugging the trace (\ref{eq:spin-g}) into Eq.~(\ref{effectiveaction-g}), we have 
\beq
{\cal L}_g^{(1')}= -  i \int_0^\infty \frac{ds}{s}\,  {\rm e}^{-\epsilon s} 
 \left[ 
 [ \cosh(2 v \aaa s) +  \cos(2 v \bbb s) ]
 \langle x | {\rm e}^{- i \hat H_s  s}|x\rangle  - 2\langle x|{\rm e}^{-i \hat H_{s0}  s}|x\rangle 
\right] 
\, . 
\label{effectiveaction-g2}
\eeq

Using the previous result on the transition amplitude (\ref{eq:TA-x}), 
we get the coordinate-space representation of the gluon contribution 
\begin{eqnarray}
\label{effectiveaction-x-g}
{\cal L}_g^{(1')} &=& 
- \frac{1}{16\pi^2}\int_0^\infty \frac{ds}{s}\,  {\rm e}^{-\epsilon s} 
\left[\frac{ (v \aaa )(v \bbb ) [  \cosh(2 v \aaa s) +   \cos(2 v \bbb s) ]   }{ \sinh (v \aaa s) \sin (v \bbb s)} -  \frac{2}{s^2} \right] 
\, .
\end{eqnarray}
This reproduces the known result \cite{Batalin:1976uv, Savvidy:1977as, Yildiz:1979vv, 
Ozaki:2013sfa, Ozaki:2015yja} (see also Ref.~\cite{Savvidy:2019grj} for a recent review paper). 
Likewise, using the previous result in Eq.~(\ref{eq:ansatz}), 
we get the momentum-space representation of the gluon contribution 
\begin{eqnarray}
\label{effectiveaction-p-g}
{\cal L}_g^{(1')} &=& 
- i \int_0^\infty \frac{ds}{s}\,  {\rm e}^{- \epsilon s} 
\pint
\left[
\frac{   \cosh(2 v \aaa s) +  \cos(2 v \bbb s)   } { \cosh (v \aaa s) \cos (v \bbb s) } 
\e^{  i \frac{  \tanh(v \aaa s)}{ v \aaa } p_\para^2  + i \frac{   \tan(  v \bbb s)}{v \bbb} p_\perp^2  } 
- 2  \e^{ i p^2 s}
\right] 
\, .
\nn
\\
\end{eqnarray}

Now, we take the sum of the gluon and ghost contributions which we denote as 
\begin{eqnarray}
{\cal L}_{\rm YM} = {\cal L}_\gluon + {\cal L}_\ghost
=
\sum_{a=1}^8 \left [ \,  {\cal L}_L^{(1^\prime)} + {\cal L}_T^{(1^\prime)}   \, \right]
\, .
\end{eqnarray}
The rightmost side is given by the above one-loop results 
\begin{subequations}
\label{eq:YM-LT}
\begin{eqnarray}
\label{eq:YM-L}
{\cal L}_L^{(1^\prime)}  &=&
- i \! \! \int_0^\infty \! \frac{ds}{s}  {\rm e}^{- \epsilon s}\! \!\!
\pint \frac{  \cosh(2 v \aaa s) } { \cosh (v \aaa s) \cos (v \bbb s) } 
\e^{  i \frac{  \tanh(v \aaa s)}{ v \aaa } p_\para^2  + i \frac{   \tan(  v \bbb s)}{v \bbb} p_\perp^2  } 
-  \left. {\cal L}_s^{(1^\prime)}  \right|_{q_f b \to v \bbb}, 
\\
\label{eq:YM-T}
 {\cal L}_T^{(1^\prime)} &=&
- i \! \! \int_0^\infty \! \frac{ds}{s} {\rm e}^{- \epsilon s} \!\!\!
\pint \frac{    \cos(2 v \bbb s)   } { \cosh (v \aaa s) \cos (v \bbb s) } 
\e^{  i \frac{  \tanh(v \aaa s)}{ v \aaa } p_\para^2  + i \frac{   \tan(  v \bbb s)}{v \bbb} p_\perp^2  } 
\, .
\end{eqnarray}
\end{subequations}
The second term in Eq.~(\ref{eq:YM-L}) is the ghost contribution (\ref{eq:HE-YM}) with a negative sign. 
The meaning of the subscripts $ L $ and $ T $ will become clear shortly.

As we have done for QED, we apply the Landau-level decomposition to 
the momentum-space representations (\ref{eq:YM-LT}) that contain the following momentum integrals 
\begin{subequations}
\label{eq:I_perp-g}
\begin{eqnarray}
I_{L} (\aaa,\bbb) &:=& 
 \int \frac{d^4p}{(2\pi)^4} 
\frac{  \cosh(2 v \aaa s)    } { \cosh (v \aaa s) \cos (v \bbb s) } 
\e^{   i \frac{  \tanh(v \aaa s)}{ v \aaa } p_\para^2  + i \frac{   \tan(  v \bbb s)}{v \bbb} p_\perp^2  } 
\nn
\\
&=& \left[  \frac{  i}{(2\pi)^2} \frac{\pi v \aaa  }{ i \sinh(v \aaa s)} \right]
\cosh(2 v \aaa s)   \int \frac{d^2p_\perp}{(2\pi)^2} 
\frac{  1 } { \cos (v \bbb s) }  \e^{   i \frac{   \tan( v \bbb s)}{v\bbb} p_\perp^2  }
\, ,
\\
I_{T} (\aaa,\bbb) &:=&  
 \int \frac{d^4p}{(2\pi)^4} 
\frac{  \cos(2 v \bbb s)   } { \cosh (v \aaa s) \cos (v \bbb s) } 
\e^{   i \frac{  \tanh(v \aaa s)}{ v \aaa } p_\para^2  + i \frac{   \tan(  v \bbb s)}{v \bbb} p_\perp^2  } 
\nn
\\
&=& \left[  \frac{  i}{(2\pi)^2} \frac{\pi v \aaa }{ i \sinh(v \aaa s)} \right]
\int \frac{d^2p_\perp}{(2\pi)^2} 
\frac{    \cos(2 v \bbb s)   } {\cos (v \bbb s) }  \e^{   i \frac{   \tan( v \bbb s)}{v\bbb} p_\perp^2  }
\, .
\end{eqnarray}
\end{subequations}
We have performed the Gaussian integrals. 
As in the previous section, we use Eq.~(\ref{eq:g_Laguerre}) with the replacement, $ q_f b\to  v\bbb $. 
The $ I_{L} (\aaa,\bbb)  $ has the same structure as its counterpart in scalar QED, 
while the $ I_{T} (\aaa,\bbb)  $ has an additional factor of $  \cos(2v \bbb s)  $. 
Note that  $ \cos(2v \bbb s) = (- z^{-1} - z)/2  $, which will shift the powers of $  z$ 
and thus the energy levels by one unit. 
Accordingly, the $\bbb$-dependent parts are decomposed as 
\begin{subequations}
\begin{eqnarray}
I_{L} (\aaa,\bbb) &=& 
 \left[  \frac{  i}{(2\pi)^2} \frac{\pi v \aaa  }{ i \sinh(v \aaa s)} \right] \cosh(2 v \aaa s)  
 \times  \left[ \frac{|v \bbb|}{2\pi} \right]  \sum_{n=0}^{\infty} \e^{ -i (2n+1) |v\bbb|  s}  
\, ,
\\
I_{T} (\aaa,\bbb) &=& 
 \left[  \frac{  i}{(2\pi)^2} \frac{\pi v \aaa  }{ i \sinh(v \aaa s)} \right]
 \times \frac{1}{2} \left[ \frac{|v \bbb|}{2\pi} \right]  \sum_{n=0}^{\infty} 
\left[ \e^{- i  (2n-1) |v\bbb|  s}   +  \e^{ -i (2n+3) |v\bbb|  s}   \right]
\, ,
\end{eqnarray}
\end{subequations}
where the $ p_\perp $ integrals have been performed as in scalar QED (cf. Appendix~\ref{sec:I_perp}).

Plugging those results back to Eq.~(\ref{eq:YM-LT}), 
we obtain the HE effective Lagrangian for the Yang-Mills theory 
\begin{subequations}
\label{eq:HE-Landau-YM}
\begin{eqnarray}
&&
{\cal L}_L^{(1^\prime)}  =
\left[  \frac{| v\bbb|} {2\pi} \right] \sum_{n=0}^\infty
\left[ - \frac{i}{2\pi}\int_0^\infty \frac{ds}{s} \e^{- \epsilon s} \e^{-i  (m_{n}^{v})^2 s}   (v \aaa) \sinh(v \aaa s)  \right]
\, , 
\label{HE-Landau-L}
\\
&&
{\cal L}_T^{(1^\prime)}  =
\left[  \frac{| v\bbb|} {2\pi} \right] \sum_{n=0}^\infty 
\left[ - \frac{i}{8\pi}\int_0^\infty \frac{ds}{s} \e^{- \epsilon s} 
\left( \e^{-i (m_{n-1}^{v})^2  s} + \e^{-i  (m_{n+1}^{v})^2 s}  \right)  \frac{ v \aaa }{ \sinh(v \aaa s) } \right]
\, .
\label{HE-Landau-T}
\end{eqnarray}
\end{subequations}
As in the scalar QED result (\ref{HE-Landau-s}), we can drop the $ \epsilon $ parameter in $ {\cal L}_T^{(1^\prime)}  $, 
because the proper-time integral at each $  n$ does not have singularities on the real axis and is convergent, 
except for the divergence at the origin which is common to the free theory. 
In the above expressions, we have defined the effective mass 
\begin{eqnarray}
(m_n^v)^2 = (2n+1) |v \bbb|
\, .
\end{eqnarray}
Of course, perturbative gluons do not have a mass gap, i.e., $\lim_{\bbb\to0} m_n^v =0  $. 
The lowest energy spectrum in $ {\cal L}_L^{(1^\prime)} $ is $ (m_0^v)^2 =  |v \bbb| $, 
while $ {\cal L}_T^{(1^\prime)} $ contains two series of the Landau levels 
starting at $ (m_{-1}^v)^2 = - |v \bbb| $ and $ (m_1^v)^2 = + 3|v \bbb| $. 
The difference among them originates from the Zeeman splitting for the vector boson (cf. Fig.~\ref{fig:spectrum}). 
According to this observation, the former and latter two modes are identified 
with the longitudinal and two transverse modes, respectively. 
Without an electric field ($ \aaa = 0 $), the longitudinal-mode contribution vanishes, i.e., $ \lim_{\aaa \to 0} {\cal L}_L^{(1^\prime)}   =0  $. 
Each transverse-mode contribution is decomposed into the Landau degeneracy factor times the (1+1)-dimensional 
HE Lagrangian for a spinless particle [cf. the scalar QED result (\ref{HE-Landau-s})]. 
Both the transverse gluons and the complex scalar field have two degrees of freedom. 
Note that the ground state of one of the transverse modes is tachyonic, i.e., $ (m^v_{-1})^2 = - |v\bbb| $. 
The appearance of this mode is known as the Nielsen-Olesen instability~\cite{Nielsen:1978rm}. 



Similar to the discussion around Eq.~(\ref{eq:trigo}), one can most conveniently get the Landau-level representation 
starting from the standard form (\ref{effectiveaction-x-g}). 
Together with Eq.~(\ref{eq:trigo}), one can apply an expansion 
\begin{eqnarray}
\label{eq:trigo2}
\cos(2x) \sec x 
= i \frac{ \, \e^{ix} + \e^{-3ix} \, }{1-\e^{-2i  x} }  
= i \sum_{n=0}^\infty \left(  \e^{-i (2n-1) x}  +  \e^{-i (2n +3) x}  \right)
\, .
\end{eqnarray}
Those two terms yield the spectra of the transverse modes resultant 
from the Landau-level discretization and the Zeeman effect, 
while the other one term yields the spectrum of the longitudinal mode.

\subsection{Gluonic Schwinger mechanism}

Here, we discuss the imaginary part of the Yang-Mills part $ {\cal L}_{\rm YM} $. 
We find two special features regarding the gluon dynamics. 
First, notice that the integrand for the longitudinal mode $ {\cal L}_L^{(1^\prime)}   $ 
is regular everywhere in the complex plane. 
Therefore, there is no possible source of an imaginary part there. 
This means that the longitudinal modes are not produced 
by the Schwinger mechanism as on-shell degrees of freedom (see Ref.~\cite{Ambjorn:1982bp} 
for the detailed discussions from the perspective of the canonical quantization 
and the method of the Bogoliubov transformation). 

Second, the ground state of the transverse mode is tachyonic as mentioned above. 
Namely, the spectrum is given as $ ( m_{n-1}^v)^2 =  - |v \aaa|$. 
Picking up this contribution in the series of the Landau levels, we have 
\begin{eqnarray}
{\cal L}_{\rm NO}^{(1^\prime)}  =
\left[  \frac{| v\bbb|} {2\pi} \right]
\left[ - \frac{i}{8\pi}\int_0^\infty \frac{ds}{s}  \e^{ i |v \bbb| s}  \frac{ v \aaa }{ \sinh(v \aaa s) } \right]
\, .
\label{HE-Landau-g-NO}
\end{eqnarray}
With the tachyonic dispersion relation, the proper-time integral seems not to converge in the lower half plane 
on first sight. 
Therefore, one possible way of computing the integral is to close the contour in the upper half plane. 
Collecting the residues at $ s = i s_\sigma \ (\sigma \geq 1) $ on the positive imaginary axis, 
we obtain 
\begin{eqnarray}
\Im m {\cal L}_{\rm NO}^{(1^\prime)} = \left[  \frac{|v \bbb|} {2\pi} \right]
 \sum_{\sigma=1}^\infty (-1)^{\sigma-1} \frac{ |v \aaa| }{8\pi \sigma}   \e^{-  \left| \frac{  \bbb }{ \aaa} \right| \pi \sigma } 
\, .
 \label{imaginary_E-Landau-g-NO}
\end{eqnarray}
This result has the same form as the scalar QED result (\ref{imaginary_E-Landau-s}) 
up to the replacement of the effective mass by $   |v\bbb|  $. 
On the other hand, if we first assume a positivity $ ( m_{-1}^v)^2 > 0 $ and rotate the contour in the lower half plane, 
we find 
\begin{eqnarray}
\Im m {\cal L}_{\rm NO}^{(1^\prime)} &=& \left[  \frac{|v \bbb|} {2\pi} \right]
 \sum_{\sigma=1}^\infty (-1)^{\sigma-1} \frac{ |v \aaa| }{8\pi \sigma}   \e^{-  ( m_{-1}^v)^2 s_\sigma} 
 \nn
 \\
&\to&
 \left[  \frac{|v \bbb|} {2\pi} \right]
 \sum_{\sigma=1}^\infty (-1)^{\sigma-1} \frac{ |v \aaa| }{8\pi \sigma}   \e^{ \left| \frac{  \bbb }{ \aaa} \right| \pi \sigma}  
\, .
 \label{imaginary_E-Landau-g-NO-2}
\end{eqnarray}
In the second line, we have performed an analytic continuation to the negative region $ ( m_{-1}^v)^2 < 0 $. 
In this result, the imaginary part grows exponentially with an increasing chromo-magnetic field $ |\bbb| $ 
and suggests that the vacuum persistence probability decreases drastically. 
This result seems to us more physically sensible than the exponentially suppressed result (\ref{imaginary_E-Landau-g-NO}) 
since the presence of the tachyonic mode may imply an instability of the perturbative vacuum. 
However, we do not have a clear mathematical reason why the latter should provide the correct result. 
This point is still an open question. 
A consistent result has been obtained in one of preceding studies for the pair production rate 
from the method of the canonical quantization \cite{Tanji:2011di}. 

It may be worth mentioning that the treatment of the Nielsen-Olesen instability 
(without a chromo-electric field) has been controversial for quite some time \cite{
Nielsen:1978rm, Yildiz:1979vv, Claudson:1980yz, Leutwyler:1980ma, Schanbacher:1980vq, 
Ambjorn:1982nd,  Ambjorn:1983ne, Dittrich:1983ej, Elizalde:1984zv} (see Ref.~\cite{Savvidy:2019grj} for a recent review).
We are not aware of a clear answer to either case with or without a chromo-electric field. 
The correspondences between the relevant physical circumstances (or boundary conditions) 
and the contours of the proper-time integral may need to be clarified 
(see Refs.~\cite{Schanbacher:1980vq, Ambjorn:1982nd, Ambjorn:1983ne, Elizalde:1984zv} and 
somewhat related studies on the fate of the chiral condensate in electric fields \cite{Cohen:2007bt, Copinger:2018ftr}).


The imaginary parts from the other Landau levels can be obtained 
by enclosing the contour in the lower half plane as before. 
Summing all the contributions, we obtain the total imaginary part of the Yang-Mills contribution 
\begin{eqnarray}
\Im m {\cal L}_g^{(1^\prime)}  &=& \Im m {\cal L}_T^{(1^\prime)}  
 \label{imaginary_E-Landau-g}
\\
&=&
\Im m {\cal L}_{\rm NO}^{(1^\prime)} 
+  \left[  \frac{|v \bbb|} {2\pi} \right]  \sum_{\sigma=1}^\infty (-1)^{\sigma-1}  \frac{ |v \aaa| }{8\pi \sigma} 
\left[ \e^{- (m_{1}^{v})^2 s_\sigma}
+ \sum_{n=1}^\infty  \left( \, \e^{- (m_{n-1}^{v})^2 s_\sigma} + \e^{-  (m_{n+1}^{v})^2 s_\sigma}  \, \right)
 \right]
\, .
\nn
\end{eqnarray}
The Nielsen-Olesen mode discussed above is isolated in the first term. 
The critical field strengths for the other modes should read  
\begin{eqnarray}
 {\cal E}^c_n \equiv \frac{( m_{n}^{v})^2}{| v |} = (2n+1) |  \bbb| \quad (n \geq 0)
 \, .
\end{eqnarray}
By using the critical field strength, the imaginary part is represented as 
\begin{eqnarray}
\Im m {\cal L}_g^{(1^\prime)} 
= &&
\Im m {\cal L}_{\rm NO}^{(1^\prime)} 
+
 \left[  \frac{|v \bbb|} {2\pi} \right]  \sum_{\sigma=1}^\infty (-1)^{\sigma-1}  \frac{ |v \aaa| }{8\pi \sigma} 
 \left[  \e^{- \frac{ {\cal E}^c_{0} }{\aaa} \pi\sigma  }
 +
2 \sum_{n=1}^\infty   \e^{- \frac{ {\cal E}^c_{n} }{\aaa} \pi\sigma  } 
 \right]
 \, .
\end{eqnarray}
Note that the exponential factors do not depend on the coupling constant 
because of the cancellation in the absence of a mass term. 
Having rearranged the Landau-level summation, 
we now clearly see the two-fold spin degeneracies in the higher states (cf. Fig.~\ref{fig:spectrum}).

\section{Summary}

In the HE effective action at the one-loop level, 
we found a complete factorization of the transverse and longitudinal parts 
with respect to the direction of the magnetic field. 
Furthermore, we have shown the analytic results 
in the form of the summation over the all-order Landau levels, 
and identified the differences among the scalar particles, fermions, and gluons 
on the basis of the Zeeman energies which depend on the spin size.

Based on the Landau-level representations, 
we discussed the Schwinger mechanism in the coexistent electric and magnetic fields. 
The Schwinger mechanism is enhanced by the lowest-Landau-level contribution 
in spinor QED thanks to the Landau degeneracy factor and the fact that 
the ground-state energy is independent of the magnetic field strength. 
In contrast, the Schwinger mechanism is suppressed in scalar QED due to a stronger exponential suppression 
because the ground-state energy increases in a magnetic field due to the absence of the Zeeman shift, 
although the Landau degeneracy factor is still there. 
For the gluon production in the cavariantly constant chromo-electromagnetic field, 
we explicitly showed the cancellation between the longitudinal-gluon and ghost contributions 
identified on the basis of the Zeeman energy. 
The ground-state transverse mode is also explicitly identified with the Nielsen-Olesen instability mode. 
The presence of the instability mode may induce the exponential growth of 
the imaginary part (\ref{imaginary_E-Landau-g-NO-2}) (cf. Ref.~\cite{Tanji:2011di}). 
Nevertheless, a clear mathematical verification of Eq.~(\ref{imaginary_E-Landau-g-NO-2}), 
against Eq.~(\ref{imaginary_E-Landau-g-NO}) with the exponential suppression, 
is left as an open problem.

Extensions to finite temperature/density \cite{Dittrich:1979ux, Muller:1980kf, Dittrich:1980nh, Gies:1998vt, Gies:1999vb}
and higher-loop diagrams \cite{Ritus:1998jm, Dunne:2004nc, Gies:2016yaa, Huet:2017ydx} 
are also left as interesting future works. 
While there is no interlevel mixing at the one-loop level, 
we would expect the occurrence of interlevel transitions 
via interactions with the dynamical gauge fields in the higher-loop diagrams.

%
%
%
%

\vspace{0.5cm}

{\it Note added}.---In completion of this work, 
the authors noticed a new paper \cite{Karbstein:2019oej} 
in which the expansion method (\ref{eq:trigo}) was applied to 
the imaginary part of the effective action for spinor QED.

\vspace{0.5cm}

{\it Acknowledgments}.--- 
The authors thank Yoshimasa Hidaka for discussions.

\appendix

\section{Spinor trace in external fields}

\label{sec:traces}

Here, we compute the Dirac-spinor trace of the exponential factor 
\begin{eqnarray}
\tr \left[ \e^{ - i\frac{q_f}{2}s F\sigma } \right] =
 \sum_{n=0}^{\infty} \frac{1}{(2n)!} \left( - i\frac{q_f}{2}s \right)^{2n}  \tr \left[ (F\sigma)^{2n}  \right] 
\label{eq:app1}
\, . 
\end{eqnarray}
In the above expansion, we used the fact that the trace of the odd-power terms vanish 
\begin{eqnarray}
\tr \left[ (F\sigma)^{2n+1} \right] = 0 
\, .
\end{eqnarray}
This is because there is no scalar combinations composed of odd numbers of $ F^{\mu\nu} $. 
On the other hand, the even-power terms may be given as functions of $ \F $ and $\G  $. 
Indeed, one can explicitly show this fact with the following identifies 
\begin{subequations}
\begin{eqnarray}
&&
\left\{ \sigma^{\mu\nu} , \sigma^{\alpha\beta} \right\} = 2 \left(
g^{\mu\alpha} g^{\nu\beta} - g^{\mu\beta} g^{\nu\alpha} + i \gamma^5 \epsilon^{\mu\nu\alpha\beta} \right ) 
\, , 
\\
&&
(F\sigma)^2 = \frac{1}{2} F_{\mu\nu} F_{\alpha\beta} \left\{ \sigma^{\mu\nu} , \sigma^{\alpha\beta} \right\}
 = 8 \left( \F + i \gamma^5 \G \right)
 \, .
\end{eqnarray}
\end{subequations}
Since the $\gamma^5$ takes eigenvalues $\pm 1$, 
the diagonalized form is given by 
\begin{eqnarray}
 \tr \left[ (F\sigma)^{2n}  \right] = 
 2^{2n}  \tr \left[ {\rm diag} \bigg( \,  (i a + b) ,\,  - (i a + b), \,  (i a - b), \,  - (i a - b) \, \bigg) ^{2n}  \right] 
\, ,
\end{eqnarray}
where 
we have rewritten the combinations of $ \F, \ \G  $ by the Poirncar\'e invariants: 
\begin{eqnarray}
\sqrt{ \F \pm i \G } 
=  \sqrt{ \frac{1}{2} (b^2-a^2) \mp i ab }  
=\pm \frac{1}{\sqrt{2}} ( i a \mp b )
\, .
\end{eqnarray}
By taking the trace in the diagonalizing basis, we find 
\begin{eqnarray}
\tr \left[ \e^{ - i\frac{q_f}{2}s F\sigma } \right] &=& 2 \left[ \, \cos( q_f(ia +b )s) + \cos( q_f(ia-b)s) \, \right] \nonumber\\
&=& 4 \cosh( q_f  a s ) \cos(q_f b s) 
\label{eq:eq1}
\, .
\end{eqnarray}

%

\section{Transverse-momentum integrals}

\label{sec:I_perp}


We perform the following two types of the integrals 
\begin{subequations}
\begin{eqnarray}
&&
\label{eq:I-s0}
2 (-1)^n \int \frac{d^2p_\perp}{(2\pi)^2} \e^{- \frac{ u_\perp }{ 2 } }  L_n (u_\perp)
=  \frac{|q_f b|}{2\pi} \frac{(-1)^n}{2} \int du_\perp \e^{- \frac{ u_\perp }{ 2 } } L_n (u_\perp)
\, ,
\\
&&
\label{eq:I-f0}
2 (-1)^n \int \frac{d^2p_\perp}{(2\pi)^2} \e^{- \frac{ u_\perp }{ 2 } }  L_n^{-1} (u_\perp)
= \frac{|q_f b|}{2\pi} \frac{(-1)^n}{2}  \int du_\perp \e^{- \frac{ u_\perp }{ 2 } } L_n^{-1} (u_\perp)
\, .
\end{eqnarray}
\end{subequations}
The first and second ones appear in the bosonic (as well as ghost) and fermionic contributions, respectively. 
Actually, they are connected by the recursive relation for the Laguerre polynomials, 
and we can avoid repeating the similar computations. 

%

We first perform the integral for the bosonic one: 
\begin{eqnarray}
I_n^s &:=& \int_0^\infty d\zeta \e^{ -  \zeta/2 }  L_{n} \left(  \zeta  \right)
\, .
\end{eqnarray}
By the use of a derivative formula $ d L_{n+1}^\alpha(\zeta)/d\zeta = - L_{n}^{\alpha+1}(\zeta) $ 
for $n\geq  0 $ \cite{AssociatedLaguerrePolynomial}, 
we find  
\begin{eqnarray}
I_n^s &=& - \int_0^\infty d\zeta \e^{ -  \zeta/2 }  \frac{ d L_{n+1}^{-1}\left(  \zeta  \right) }{ d \zeta}
\nn
\\
&=& - [ \e^{ -  \zeta/2 }  L_{n+1}^{-1} (  \zeta )  ]_0^\infty
- \frac{1}{2} \int_0^\infty d\zeta \e^{ -  \zeta/2 }  L_{n+1}^{-1}\left(  \zeta  \right)
\nn
\\
&=&- \frac{1}{2} \int_0^\infty d\zeta \e^{ -  \zeta/2 } [ L_{n+1} (\zeta)-  L_{n}  \left(  \zeta  \right)]
\, ,
\end{eqnarray}
where the surface term vanishes. 
To reach the last line, we applied the recursive relation 
$ L_{n+1}^{\alpha-1} (\zeta)=  L_{n+1}^{\alpha}  (\zeta) -  L_{n}^{\alpha}  (\zeta) $. 
The above relation means that 
\begin{eqnarray}
I_{n+1}^s = - I_n^s
\, .
\end{eqnarray}
Since $ L_0^\alpha (\zeta)=1 $ for any $  \alpha$ and $ \zeta $, 
we can easily get $ I_0^s = 2 $. 
Therefore, we reach a simple result 
\begin{eqnarray}
I_n^s = 2 (-1)^n
\, .
\end{eqnarray}

For the fermion contribution, we need to perform the integral 
\begin{eqnarray}
I_n^f := \int_0^\infty d\zeta \e^{ -  \zeta/2 }  L_{n}^{-1}\left(  \zeta  \right)
\, .
\end{eqnarray}
By applying the above recursive relation for $ n \geq 1 $, 
we immediately get a connection between the fermionic and bosonic ones: 
\begin{eqnarray}
I_n^f =  I_n^s - I_{n-1}^s = 2  I_n^s 
\, .
\end{eqnarray}
When $ n =0 $, we can separately perform the integral to get $I_0^f = 2   $. 
Therefore, we get 
\begin{eqnarray}
I_n^f = 2 \kappa_n (-1)^n
\, ,
\end{eqnarray}
with $ \kappa_n = 2 - \delta_{n 0} $ for $ n\geq0 $. 
Summarizing above, we have obtained the analytic results for the integrals 
\begin{subequations}
\begin{eqnarray}
&&
\label{eq:I-s}
2 (-1)^n \int \frac{d^2p_\perp}{(2\pi)^2} \e^{- \frac{ u_\perp }{ 2 } }  L_n (u_\perp)
=  \frac{|q_f b|}{2\pi} 
\, ,
\\
&&
\label{eq:I-f}
2 (-1)^n \int \frac{d^2p_\perp}{(2\pi)^2} \e^{- \frac{ u_\perp }{ 2 } }  L_n^{-1} (u_\perp)
= \kappa_n \frac{|q_f b|}{2\pi} 
\, .
\end{eqnarray}
\end{subequations}

\section{Covariantly constant chromo field}

\label{sec:covariantly-constant}

To find a solution to Eq.~(\ref{eq:CCF}), 
we evaluate a quantity $[ D_\lambda, D_\sigma] ^{ab} \nonF_{\mu\nu}^b$ in two ways. 
First, the above condition immediately leads to $  [ D_\lambda, D_\sigma] ^{ab} \nonF_{\mu\nu}^b = 0$. 
On the other hand, the commutator can be written by the field strength tensor 
and the structure constant. 
Therefore, the covariantly constant field satisfies a condition 
\begin{eqnarray}
f^{abc} \nonF_{\mu\nu}^b \nonF_{\lambda \sigma}^c = 0
\label{eq:fGG}
\,  .
\end{eqnarray}
Since the four Lorentz indices are arbitrary, the above condition is satisfied only when 
the contractions of the color indices vanish. 
Therefore, we find the solution in a factorized form (\ref{eq:fact}).

For $N_c=3$, the effective color charges $ w_k $ have the three components 
\begin{eqnarray}
w_k = \frac{ g }{ \sqrt{3} }  \sin \left ( \theta + (2k-1) \frac{\pi}{3} \right )
\, , \ \ \ k=1,2,3
\label{eq:angle-fund-rep}
\, ,
\end{eqnarray}
while those for the adjoint representation $  v^{a}  $ have six non-vanishing components  
\begin{eqnarray}
\begin{array}{ll}
\displaystyle
v^a = g \sin \left ( \theta_{\rm ad} + (2a-1) \frac{\pi}{3} \right ) \, , & a=1,2,3 \, , \\
\displaystyle
v^a = - g \sin \left ( \theta_{\rm ad} + (2a-1) \frac{\pi}{3} \right ) \, , & a=5,6,7 \, , \\
\displaystyle
v^a = 0 \, , & a=4,8 \, .
\end{array}
\label{eq:angle-adj-rep}
\end{eqnarray}
The color directions $ \theta $ and $ \theta_{\rm ad} $ are specified 
by the second Casimir invariant \cite{Batalin:1976uv, Yildiz:1979vv, Ambjorn:1982bp, 
Gyulassy:1986jq, Suganuma:1991ha, Nayak:2005yv, Nayak:2005pf, Tanji:2010eu, Ozaki:2013sfa, Ozaki:2015yja}.

\bibliography{bib-Landau}

\begin{thebibliography}{83}%
\makeatletter
\providecommand \@ifxundefined [1]{%
 \@ifx{#1\undefined}
}%
\providecommand \@ifnum [1]{%
 \ifnum #1\expandafter \@firstoftwo
 \else \expandafter \@secondoftwo
 \fi
}%
\providecommand \@ifx [1]{%
 \ifx #1\expandafter \@firstoftwo
 \else \expandafter \@secondoftwo
 \fi
}%
\providecommand \natexlab [1]{#1}%
\providecommand \enquote  [1]{``#1''}%
\providecommand \bibnamefont  [1]{#1}%
\providecommand \bibfnamefont [1]{#1}%
\providecommand \citenamefont [1]{#1}%
\providecommand \href@noop [0]{\@secondoftwo}%
\providecommand \href [0]{\begingroup \@sanitize@url \@href}%
\providecommand \@href[1]{\@@startlink{#1}\@@href}%
\providecommand \@@href[1]{\endgroup#1\@@endlink}%
\providecommand \@sanitize@url [0]{\catcode `\\12\catcode `\$12\catcode
  `\&12\catcode `\#12\catcode `\^12\catcode `\_12\catcode `\%12\relax}%
\providecommand \@@startlink[1]{}%
\providecommand \@@endlink[0]{}%
\providecommand \url  [0]{\begingroup\@sanitize@url \@url }%
\providecommand \@url [1]{\endgroup\@href {#1}{\urlprefix }}%
\providecommand \urlprefix  [0]{URL }%
\providecommand \Eprint [0]{\href }%
\providecommand \doibase [0]{http://dx.doi.org/}%
\providecommand \selectlanguage [0]{\@gobble}%
\providecommand \bibinfo  [0]{\@secondoftwo}%
\providecommand \bibfield  [0]{\@secondoftwo}%
\providecommand \translation [1]{[#1]}%
\providecommand \BibitemOpen [0]{}%
\providecommand \bibitemStop [0]{}%
\providecommand \bibitemNoStop [0]{.\EOS\space}%
\providecommand \EOS [0]{\spacefactor3000\relax}%
\providecommand \BibitemShut  [1]{\csname bibitem#1\endcsname}%
\let\auto@bib@innerbib\@empty
\bibitem [{\citenamefont {Heisenberg}\ and\ \citenamefont
  {Euler}(1936)}]{Heisenberg:1935qt}%
  \BibitemOpen
  \bibfield  {author} {\bibinfo {author} {\bibfnamefont {W.}~\bibnamefont
  {Heisenberg}}\ and\ \bibinfo {author} {\bibfnamefont {H.}~\bibnamefont
  {Euler}},\ }\bibfield  {title} {\enquote {\bibinfo {title} {{Consequences of
  Dirac's theory of positrons}},}\ }\href {\doibase 10.1007/BF01343663}
  {\bibfield  {journal} {\bibinfo  {journal} {Z. Phys.}\ }\textbf {\bibinfo
  {volume} {98}},\ \bibinfo {pages} {714--732} (\bibinfo {year} {1936})},\
  \Eprint {http://arxiv.org/abs/physics/0605038} {arXiv:physics/0605038
  [physics]} \BibitemShut {NoStop}%
\bibitem [{\citenamefont {Schwinger}(1951)}]{Schwinger:1951nm}%
  \BibitemOpen
  \bibfield  {author} {\bibinfo {author} {\bibfnamefont {Julian~S.}\
  \bibnamefont {Schwinger}},\ }\bibfield  {title} {\enquote {\bibinfo {title}
  {{On gauge invariance and vacuum polarization}},}\ }\href {\doibase
  10.1103/PhysRev.82.664} {\bibfield  {journal} {\bibinfo  {journal} {Phys.
  Rev.}\ }\textbf {\bibinfo {volume} {82}},\ \bibinfo {pages} {664--679}
  (\bibinfo {year} {1951})}\BibitemShut {NoStop}%
\bibitem [{\citenamefont {Nambu}(1950)}]{Nambu:1950rs}%
  \BibitemOpen
  \bibfield  {author} {\bibinfo {author} {\bibfnamefont {Yoichiro}\
  \bibnamefont {Nambu}},\ }\bibfield  {title} {\enquote {\bibinfo {title} {{The
  use of the Proper Time in Quantum Electrodynamics}},}\ }\href {\doibase
  10.1143/PTP.5.82} {\bibfield  {journal} {\bibinfo  {journal} {Prog. Theor.
  Phys.}\ }\textbf {\bibinfo {volume} {5}},\ \bibinfo {pages} {82--94}
  (\bibinfo {year} {1950})}\BibitemShut {NoStop}%
\bibitem [{\citenamefont {Feynman}(1950)}]{Feynman:1950ir}%
  \BibitemOpen
  \bibfield  {author} {\bibinfo {author} {\bibfnamefont {R.~P.}\ \bibnamefont
  {Feynman}},\ }\bibfield  {title} {\enquote {\bibinfo {title} {{Mathematical
  formulation of the quantum theory of electromagnetic interaction}},}\ }\href
  {\doibase 10.1103/PhysRev.80.440} {\bibfield  {journal} {\bibinfo  {journal}
  {Phys. Rev.}\ }\textbf {\bibinfo {volume} {80}},\ \bibinfo {pages} {440--457}
  (\bibinfo {year} {1950})}\BibitemShut {NoStop}%
\bibitem [{\citenamefont {Fock}(1937)}]{Fock:1937dy}%
  \BibitemOpen
  \bibfield  {author} {\bibinfo {author} {\bibfnamefont {V.}~\bibnamefont
  {Fock}},\ }\bibfield  {title} {\enquote {\bibinfo {title} {{Proper time in
  classical and quantum mechanics}},}\ }\href@noop {} {\bibfield  {journal}
  {\bibinfo  {journal} {Phys. Z. Sowjetunion}\ }\textbf {\bibinfo {volume}
  {12}},\ \bibinfo {pages} {404--425} (\bibinfo {year} {1937})}\BibitemShut
  {NoStop}%
\bibitem [{\citenamefont {Sauter}(1931)}]{Sauter:1931zz}%
  \BibitemOpen
  \bibfield  {author} {\bibinfo {author} {\bibfnamefont {Fritz}\ \bibnamefont
  {Sauter}},\ }\bibfield  {title} {\enquote {\bibinfo {title} {{Uber das
  Verhalten eines Elektrons im homogenen elektrischen Feld nach der
  relativistischen Theorie Diracs}},}\ }\href {\doibase 10.1007/BF01339461}
  {\bibfield  {journal} {\bibinfo  {journal} {Z. Phys.}\ }\textbf {\bibinfo
  {volume} {69}},\ \bibinfo {pages} {742--764} (\bibinfo {year}
  {1931})}\BibitemShut {NoStop}%
\bibitem [{\citenamefont {Toll}(1952)}]{Toll:1952rq}%
  \BibitemOpen
  \bibfield  {author} {\bibinfo {author} {\bibfnamefont {John~Sampson}\
  \bibnamefont {Toll}},\ }\emph {\bibinfo {title} {{The Dispersion relation for
  light and its application to problems involving electron pairs}}},\
  \href@noop {} {Ph.D. thesis},\ \bibinfo  {school} {Princeton U.} (\bibinfo
  {year} {1952})\BibitemShut {NoStop}%
\bibitem [{\citenamefont {Klein}\ and\ \citenamefont
  {Nigam}(1964)}]{Klein:1964zza}%
  \BibitemOpen
  \bibfield  {author} {\bibinfo {author} {\bibfnamefont {James~J.}\
  \bibnamefont {Klein}}\ and\ \bibinfo {author} {\bibfnamefont {B.~P.}\
  \bibnamefont {Nigam}},\ }\bibfield  {title} {\enquote {\bibinfo {title}
  {{Dichroism of the Vacuum}},}\ }\href {\doibase 10.1103/PhysRev.136.B1540}
  {\bibfield  {journal} {\bibinfo  {journal} {Phys. Rev.}\ }\textbf {\bibinfo
  {volume} {136}},\ \bibinfo {pages} {B1540--B1542} (\bibinfo {year}
  {1964})}\BibitemShut {NoStop}%
\bibitem [{\citenamefont {Baier}\ and\ \citenamefont
  {Breitenlohner}(1967{\natexlab{a}})}]{Baier:1967zza}%
  \BibitemOpen
  \bibfield  {author} {\bibinfo {author} {\bibfnamefont {R.}~\bibnamefont
  {Baier}}\ and\ \bibinfo {author} {\bibfnamefont {P.}~\bibnamefont
  {Breitenlohner}},\ }\bibfield  {title} {\enquote {\bibinfo {title} {{Photon
  Propagation in External Fields}},}\ }\href@noop {} {\bibfield  {journal}
  {\bibinfo  {journal} {Acta Phys. Austriaca}\ }\textbf {\bibinfo {volume}
  {25}},\ \bibinfo {pages} {212--223} (\bibinfo {year}
  {1967}{\natexlab{a}})}\BibitemShut {NoStop}%
\bibitem [{\citenamefont {Baier}\ and\ \citenamefont
  {Breitenlohner}(1967{\natexlab{b}})}]{Baier:1967zzc}%
  \BibitemOpen
  \bibfield  {author} {\bibinfo {author} {\bibfnamefont {R.}~\bibnamefont
  {Baier}}\ and\ \bibinfo {author} {\bibfnamefont {P.}~\bibnamefont
  {Breitenlohner}},\ }\bibfield  {title} {\enquote {\bibinfo {title} {{The
  Vacuum refraction Index in the presence of External Fields}},}\ }\href
  {\doibase 10.1007/BF02712312} {\bibfield  {journal} {\bibinfo  {journal}
  {Nuovo Cim.}\ }\textbf {\bibinfo {volume} {B47}},\ \bibinfo {pages}
  {117--120} (\bibinfo {year} {1967}{\natexlab{b}})}\BibitemShut {NoStop}%
\bibitem [{\citenamefont {Bialynicka-Birula}\ and\ \citenamefont
  {Bialynicki-Birula}(1970)}]{BialynickaBirula:1970vy}%
  \BibitemOpen
  \bibfield  {author} {\bibinfo {author} {\bibfnamefont {Z.}~\bibnamefont
  {Bialynicka-Birula}}\ and\ \bibinfo {author} {\bibfnamefont {I.}~\bibnamefont
  {Bialynicki-Birula}},\ }\bibfield  {title} {\enquote {\bibinfo {title}
  {{Nonlinear effects in Quantum Electrodynamics. Photon propagation and photon
  splitting in an external field}},}\ }\href {\doibase 10.1103/PhysRevD.2.2341}
  {\bibfield  {journal} {\bibinfo  {journal} {Phys. Rev.}\ }\textbf {\bibinfo
  {volume} {D2}},\ \bibinfo {pages} {2341--2345} (\bibinfo {year}
  {1970})}\BibitemShut {NoStop}%
\bibitem [{\citenamefont {Brezin}\ and\ \citenamefont
  {Itzykson}(1971)}]{Brezin:1971nd}%
  \BibitemOpen
  \bibfield  {author} {\bibinfo {author} {\bibfnamefont {E.}~\bibnamefont
  {Brezin}}\ and\ \bibinfo {author} {\bibfnamefont {C.}~\bibnamefont
  {Itzykson}},\ }\bibfield  {title} {\enquote {\bibinfo {title} {{Polarization
  phenomena in vacuum nonlinear electrodynamics}},}\ }\href {\doibase
  10.1103/PhysRevD.3.618} {\bibfield  {journal} {\bibinfo  {journal} {Phys.
  Rev.}\ }\textbf {\bibinfo {volume} {D3}},\ \bibinfo {pages} {618--621}
  (\bibinfo {year} {1971})}\BibitemShut {NoStop}%
\bibitem [{\citenamefont {Adler}(1971)}]{Adler:1971wn}%
  \BibitemOpen
  \bibfield  {author} {\bibinfo {author} {\bibfnamefont {Stephen~L.}\
  \bibnamefont {Adler}},\ }\bibfield  {title} {\enquote {\bibinfo {title}
  {{Photon splitting and photon dispersion in a strong magnetic field}},}\
  }\href {\doibase 10.1016/0003-4916(71)90154-0} {\bibfield  {journal}
  {\bibinfo  {journal} {Annals Phys.}\ }\textbf {\bibinfo {volume} {67}},\
  \bibinfo {pages} {599--647} (\bibinfo {year} {1971})}\BibitemShut {NoStop}%
\bibitem [{\citenamefont {Dittrich}\ and\ \citenamefont
  {Gies}(2000)}]{Dittrich:2000zu}%
  \BibitemOpen
  \bibfield  {author} {\bibinfo {author} {\bibfnamefont {W.}~\bibnamefont
  {Dittrich}}\ and\ \bibinfo {author} {\bibfnamefont {H.}~\bibnamefont
  {Gies}},\ }\bibfield  {title} {\enquote {\bibinfo {title} {{Probing the
  quantum vacuum. Perturbative effective action approach in quantum
  electrodynamics and its application}},}\ }\href {\doibase
  10.1007/3-540-45585-X} {\bibfield  {journal} {\bibinfo  {journal} {Springer
  Tracts Mod. Phys.}\ }\textbf {\bibinfo {volume} {166}},\ \bibinfo {pages}
  {1--241} (\bibinfo {year} {2000})}\BibitemShut {NoStop}%
\bibitem [{\citenamefont {Batalin}\ \emph {et~al.}(1977)\citenamefont
  {Batalin}, \citenamefont {Matinyan},\ and\ \citenamefont
  {Savvidy}}]{Batalin:1976uv}%
  \BibitemOpen
  \bibfield  {author} {\bibinfo {author} {\bibfnamefont {I.~A.}\ \bibnamefont
  {Batalin}}, \bibinfo {author} {\bibfnamefont {Sergei~G.}\ \bibnamefont
  {Matinyan}}, \ and\ \bibinfo {author} {\bibfnamefont {G.~K.}\ \bibnamefont
  {Savvidy}},\ }\bibfield  {title} {\enquote {\bibinfo {title} {{Vacuum
  Polarization by a Source-Free Gauge Field}},}\ }\href@noop {} {\bibfield
  {journal} {\bibinfo  {journal} {Sov. J. Nucl. Phys.}\ }\textbf {\bibinfo
  {volume} {26}},\ \bibinfo {pages} {214} (\bibinfo {year} {1977})},\ \bibinfo
  {note} {[Yad. Fiz.26,407(1977)]}\BibitemShut {NoStop}%
\bibitem [{\citenamefont {Matinyan}\ and\ \citenamefont
  {Savvidy}(1978)}]{Matinyan:1976mp}%
  \BibitemOpen
  \bibfield  {author} {\bibinfo {author} {\bibfnamefont {Sergei~G.}\
  \bibnamefont {Matinyan}}\ and\ \bibinfo {author} {\bibfnamefont {G.~K.}\
  \bibnamefont {Savvidy}},\ }\bibfield  {title} {\enquote {\bibinfo {title}
  {{Vacuum Polarization Induced by the Intense Gauge Field}},}\ }\href
  {\doibase 10.1016/0550-3213(78)90463-7} {\bibfield  {journal} {\bibinfo
  {journal} {Nucl. Phys.}\ }\textbf {\bibinfo {volume} {B134}},\ \bibinfo
  {pages} {539--545} (\bibinfo {year} {1978})}\BibitemShut {NoStop}%
\bibitem [{\citenamefont {Savvidy}(1977)}]{Savvidy:1977as}%
  \BibitemOpen
  \bibfield  {author} {\bibinfo {author} {\bibfnamefont {G.~K.}\ \bibnamefont
  {Savvidy}},\ }\bibfield  {title} {\enquote {\bibinfo {title} {{Infrared
  Instability of the Vacuum State of Gauge Theories and Asymptotic Freedom}},}\
  }\href {\doibase 10.1016/0370-2693(77)90759-6} {\bibfield  {journal}
  {\bibinfo  {journal} {Phys. Lett.}\ }\textbf {\bibinfo {volume} {B71}},\
  \bibinfo {pages} {133--134} (\bibinfo {year} {1977})}\BibitemShut {NoStop}%
\bibitem [{\citenamefont {Yildiz}\ and\ \citenamefont
  {Cox}(1980)}]{Yildiz:1979vv}%
  \BibitemOpen
  \bibfield  {author} {\bibinfo {author} {\bibfnamefont {Asim}\ \bibnamefont
  {Yildiz}}\ and\ \bibinfo {author} {\bibfnamefont {Paul~H.}\ \bibnamefont
  {Cox}},\ }\bibfield  {title} {\enquote {\bibinfo {title} {{Vacuum Behavior in
  Quantum Chromodynamics}},}\ }\href {\doibase 10.1103/PhysRevD.21.1095}
  {\bibfield  {journal} {\bibinfo  {journal} {Phys. Rev.}\ }\textbf {\bibinfo
  {volume} {D21}},\ \bibinfo {pages} {1095} (\bibinfo {year}
  {1980})}\BibitemShut {NoStop}%
\bibitem [{\citenamefont {Dittrich}\ and\ \citenamefont
  {Reuter}(1983)}]{Dittrich:1983ej}%
  \BibitemOpen
  \bibfield  {author} {\bibinfo {author} {\bibfnamefont {Walter}\ \bibnamefont
  {Dittrich}}\ and\ \bibinfo {author} {\bibfnamefont {Martin}\ \bibnamefont
  {Reuter}},\ }\bibfield  {title} {\enquote {\bibinfo {title} {{Effective {QCD}
  Lagrangian With Zeta Function Regularization}},}\ }\href {\doibase
  10.1016/0370-2693(83)90268-X} {\bibfield  {journal} {\bibinfo  {journal}
  {Phys. Lett.}\ }\textbf {\bibinfo {volume} {128B}},\ \bibinfo {pages}
  {321--326} (\bibinfo {year} {1983})}\BibitemShut {NoStop}%
\bibitem [{\citenamefont {Savvidy}(2019)}]{Savvidy:2019grj}%
  \BibitemOpen
  \bibfield  {author} {\bibinfo {author} {\bibfnamefont {George}\ \bibnamefont
  {Savvidy}},\ }\bibfield  {title} {\enquote {\bibinfo {title} {{From
  Heisenberg-Euler Lagrangian to the discovery of chromomagnetic gluon
  condensation}},}\ }\href@noop {} {\  (\bibinfo {year} {2019})},\ \Eprint
  {http://arxiv.org/abs/1910.00654} {arXiv:1910.00654 [hep-th]} \BibitemShut
  {NoStop}%
\bibitem [{\citenamefont {Nielsen}\ and\ \citenamefont
  {Olesen}(1978)}]{Nielsen:1978rm}%
  \BibitemOpen
  \bibfield  {author} {\bibinfo {author} {\bibfnamefont {N.~K.}\ \bibnamefont
  {Nielsen}}\ and\ \bibinfo {author} {\bibfnamefont {P.}~\bibnamefont
  {Olesen}},\ }\bibfield  {title} {\enquote {\bibinfo {title} {{An Unstable
  Yang-Mills Field Mode}},}\ }\href {\doibase 10.1016/0550-3213(78)90377-2}
  {\bibfield  {journal} {\bibinfo  {journal} {Nucl. Phys.}\ }\textbf {\bibinfo
  {volume} {B144}},\ \bibinfo {pages} {376--396} (\bibinfo {year}
  {1978})}\BibitemShut {NoStop}%
\bibitem [{\citenamefont {Casher}\ \emph {et~al.}(1979)\citenamefont {Casher},
  \citenamefont {Neuberger},\ and\ \citenamefont {Nussinov}}]{Casher:1978wy}%
  \BibitemOpen
  \bibfield  {author} {\bibinfo {author} {\bibfnamefont {A.}~\bibnamefont
  {Casher}}, \bibinfo {author} {\bibfnamefont {H.}~\bibnamefont {Neuberger}}, \
  and\ \bibinfo {author} {\bibfnamefont {S.}~\bibnamefont {Nussinov}},\
  }\bibfield  {title} {\enquote {\bibinfo {title} {{Chromoelectric Flux Tube
  Model of Particle Production}},}\ }\href {\doibase 10.1103/PhysRevD.20.179}
  {\bibfield  {journal} {\bibinfo  {journal} {Phys. Rev.}\ }\textbf {\bibinfo
  {volume} {D20}},\ \bibinfo {pages} {179--188} (\bibinfo {year}
  {1979})}\BibitemShut {NoStop}%
\bibitem [{\citenamefont {Casher}\ \emph {et~al.}(1980)\citenamefont {Casher},
  \citenamefont {Neuberger},\ and\ \citenamefont {Nussinov}}]{Casher:1979gw}%
  \BibitemOpen
  \bibfield  {author} {\bibinfo {author} {\bibfnamefont {A.}~\bibnamefont
  {Casher}}, \bibinfo {author} {\bibfnamefont {H.}~\bibnamefont {Neuberger}}, \
  and\ \bibinfo {author} {\bibfnamefont {S.}~\bibnamefont {Nussinov}},\
  }\bibfield  {title} {\enquote {\bibinfo {title} {{MULTIPARTICLE PRODUCTION BY
  BUBBLING FLUX TUBES}},}\ }\href {\doibase 10.1103/PhysRevD.21.1966}
  {\bibfield  {journal} {\bibinfo  {journal} {Phys. Rev.}\ }\textbf {\bibinfo
  {volume} {D21}},\ \bibinfo {pages} {1966} (\bibinfo {year}
  {1980})}\BibitemShut {NoStop}%
\bibitem [{\citenamefont {Biro}\ \emph {et~al.}(1984)\citenamefont {Biro},
  \citenamefont {Nielsen},\ and\ \citenamefont {Knoll}}]{Biro:1984cf}%
  \BibitemOpen
  \bibfield  {author} {\bibinfo {author} {\bibfnamefont {T.~S.}\ \bibnamefont
  {Biro}}, \bibinfo {author} {\bibfnamefont {Holger~Bech}\ \bibnamefont
  {Nielsen}}, \ and\ \bibinfo {author} {\bibfnamefont {Joern}\ \bibnamefont
  {Knoll}},\ }\bibfield  {title} {\enquote {\bibinfo {title} {{Color Rope Model
  for Extreme Relativistic Heavy Ion Collisions}},}\ }\href {\doibase
  10.1016/0550-3213(84)90441-3} {\bibfield  {journal} {\bibinfo  {journal}
  {Nucl. Phys.}\ }\textbf {\bibinfo {volume} {B245}},\ \bibinfo {pages}
  {449--468} (\bibinfo {year} {1984})}\BibitemShut {NoStop}%
\bibitem [{\citenamefont {Kajantie}\ and\ \citenamefont
  {Matsui}(1985)}]{Kajantie:1985jh}%
  \BibitemOpen
  \bibfield  {author} {\bibinfo {author} {\bibfnamefont {K.}~\bibnamefont
  {Kajantie}}\ and\ \bibinfo {author} {\bibfnamefont {T.}~\bibnamefont
  {Matsui}},\ }\bibfield  {title} {\enquote {\bibinfo {title} {{Decay of Strong
  Color Electric Field and Thermalization in Ultrarelativistic Nucleus-Nucleus
  Collisions}},}\ }\href {\doibase 10.1016/0370-2693(85)90343-0} {\bibfield
  {journal} {\bibinfo  {journal} {Phys. Lett.}\ }\textbf {\bibinfo {volume}
  {164B}},\ \bibinfo {pages} {373--378} (\bibinfo {year} {1985})}\BibitemShut
  {NoStop}%
\bibitem [{\citenamefont {Gyulassy}\ and\ \citenamefont
  {Iwazaki}(1985)}]{Gyulassy:1986jq}%
  \BibitemOpen
  \bibfield  {author} {\bibinfo {author} {\bibfnamefont {M.}~\bibnamefont
  {Gyulassy}}\ and\ \bibinfo {author} {\bibfnamefont {A.}~\bibnamefont
  {Iwazaki}},\ }\bibfield  {title} {\enquote {\bibinfo {title} {{QUARK AND
  GLUON PAIR PRODUCTION IN SU(N) COVARIANT CONSTANT FIELDS}},}\ }\href
  {\doibase 10.1016/0370-2693(85)90711-7} {\bibfield  {journal} {\bibinfo
  {journal} {Phys. Lett.}\ }\textbf {\bibinfo {volume} {165B}},\ \bibinfo
  {pages} {157--161} (\bibinfo {year} {1985})}\BibitemShut {NoStop}%
\bibitem [{\citenamefont {Gelis}\ and\ \citenamefont
  {Tanji}(2016)}]{Gelis:2015kya}%
  \BibitemOpen
  \bibfield  {author} {\bibinfo {author} {\bibfnamefont {Francois}\
  \bibnamefont {Gelis}}\ and\ \bibinfo {author} {\bibfnamefont {Naoto}\
  \bibnamefont {Tanji}},\ }\bibfield  {title} {\enquote {\bibinfo {title}
  {{Schwinger mechanism revisited}},}\ }\href {\doibase
  10.1016/j.ppnp.2015.11.001} {\bibfield  {journal} {\bibinfo  {journal} {Prog.
  Part. Nucl. Phys.}\ }\textbf {\bibinfo {volume} {87}},\ \bibinfo {pages}
  {1--49} (\bibinfo {year} {2016})},\ \Eprint {http://arxiv.org/abs/1510.05451}
  {arXiv:1510.05451 [hep-ph]} \BibitemShut {NoStop}%
\bibitem [{\citenamefont {Ozaki}(2014)}]{Ozaki:2013sfa}%
  \BibitemOpen
  \bibfield  {author} {\bibinfo {author} {\bibfnamefont {Sho}\ \bibnamefont
  {Ozaki}},\ }\bibfield  {title} {\enquote {\bibinfo {title} {{QCD effective
  potential with strong $U(1)_{em}$ magnetic fields}},}\ }\href {\doibase
  10.1103/PhysRevD.89.054022} {\bibfield  {journal} {\bibinfo  {journal} {Phys.
  Rev.}\ }\textbf {\bibinfo {volume} {D89}},\ \bibinfo {pages} {054022}
  (\bibinfo {year} {2014})},\ \Eprint {http://arxiv.org/abs/1311.3137}
  {arXiv:1311.3137 [hep-ph]} \BibitemShut {NoStop}%
\bibitem [{\citenamefont {Bali}\ \emph {et~al.}(2013)\citenamefont {Bali},
  \citenamefont {Bruckmann}, \citenamefont {Endrodi}, \citenamefont {Gruber},\
  and\ \citenamefont {Schaefer}}]{Bali:2013esa}%
  \BibitemOpen
  \bibfield  {author} {\bibinfo {author} {\bibfnamefont {G.~S.}\ \bibnamefont
  {Bali}}, \bibinfo {author} {\bibfnamefont {F.}~\bibnamefont {Bruckmann}},
  \bibinfo {author} {\bibfnamefont {G.}~\bibnamefont {Endrodi}}, \bibinfo
  {author} {\bibfnamefont {F.}~\bibnamefont {Gruber}}, \ and\ \bibinfo {author}
  {\bibfnamefont {A.}~\bibnamefont {Schaefer}},\ }\bibfield  {title} {\enquote
  {\bibinfo {title} {{Magnetic field-induced gluonic (inverse) catalysis and
  pressure (an)isotropy in QCD}},}\ }\href {\doibase 10.1007/JHEP04(2013)130}
  {\bibfield  {journal} {\bibinfo  {journal} {JHEP}\ }\textbf {\bibinfo
  {volume} {04}},\ \bibinfo {pages} {130} (\bibinfo {year} {2013})},\ \Eprint
  {http://arxiv.org/abs/1303.1328} {arXiv:1303.1328 [hep-lat]} \BibitemShut
  {NoStop}%
\bibitem [{\citenamefont {Dittrich}\ \emph {et~al.}(1979)\citenamefont
  {Dittrich}, \citenamefont {Tsai},\ and\ \citenamefont
  {Zimmermann}}]{Dittrich:1978fc}%
  \BibitemOpen
  \bibfield  {author} {\bibinfo {author} {\bibfnamefont {Walter}\ \bibnamefont
  {Dittrich}}, \bibinfo {author} {\bibfnamefont {Wu-yang}\ \bibnamefont
  {Tsai}}, \ and\ \bibinfo {author} {\bibfnamefont {Karl-Heinz}\ \bibnamefont
  {Zimmermann}},\ }\bibfield  {title} {\enquote {\bibinfo {title} {{On the
  Evaluation of the Effective Potential in Quantum Electrodynamics}},}\ }\href
  {\doibase 10.1103/PhysRevD.19.2929} {\bibfield  {journal} {\bibinfo
  {journal} {Phys. Rev.}\ }\textbf {\bibinfo {volume} {D19}},\ \bibinfo {pages}
  {2929} (\bibinfo {year} {1979})}\BibitemShut {NoStop}%
\bibitem [{\citenamefont {Klevansky}\ and\ \citenamefont
  {Lemmer}(1989)}]{Klevansky:1989vi}%
  \BibitemOpen
  \bibfield  {author} {\bibinfo {author} {\bibfnamefont {S.~P.}\ \bibnamefont
  {Klevansky}}\ and\ \bibinfo {author} {\bibfnamefont {Richard~H.}\
  \bibnamefont {Lemmer}},\ }\bibfield  {title} {\enquote {\bibinfo {title}
  {{Chiral symmetry restoration in the Nambu-Jona-Lasinio model with a constant
  electromagnetic field}},}\ }\href {\doibase 10.1103/PhysRevD.39.3478}
  {\bibfield  {journal} {\bibinfo  {journal} {Phys. Rev.}\ }\textbf {\bibinfo
  {volume} {D39}},\ \bibinfo {pages} {3478--3489} (\bibinfo {year}
  {1989})}\BibitemShut {NoStop}%
\bibitem [{\citenamefont {Suganuma}\ and\ \citenamefont
  {Tatsumi}(1991)}]{Suganuma:1990nn}%
  \BibitemOpen
  \bibfield  {author} {\bibinfo {author} {\bibfnamefont {Hideo}\ \bibnamefont
  {Suganuma}}\ and\ \bibinfo {author} {\bibfnamefont {Toshitaka}\ \bibnamefont
  {Tatsumi}},\ }\bibfield  {title} {\enquote {\bibinfo {title} {{On the
  Behavior of Symmetry and Phase Transitions in a Strong Electromagnetic
  Field}},}\ }\href {\doibase 10.1016/0003-4916(91)90304-Q} {\bibfield
  {journal} {\bibinfo  {journal} {Annals Phys.}\ }\textbf {\bibinfo {volume}
  {208}},\ \bibinfo {pages} {470--508} (\bibinfo {year} {1991})}\BibitemShut
  {NoStop}%
\bibitem [{\citenamefont {Gusynin}\ \emph
  {et~al.}(1995{\natexlab{a}})\citenamefont {Gusynin}, \citenamefont
  {Miransky},\ and\ \citenamefont {Shovkovy}}]{Gusynin:1994xp}%
  \BibitemOpen
  \bibfield  {author} {\bibinfo {author} {\bibfnamefont {V.~P.}\ \bibnamefont
  {Gusynin}}, \bibinfo {author} {\bibfnamefont {V.~A.}\ \bibnamefont
  {Miransky}}, \ and\ \bibinfo {author} {\bibfnamefont {I.~A.}\ \bibnamefont
  {Shovkovy}},\ }\bibfield  {title} {\enquote {\bibinfo {title} {{Dimensional
  reduction and dynamical chiral symmetry breaking by a magnetic field in
  (3+1)-dimensions}},}\ }\href {\doibase 10.1016/0370-2693(95)00232-A}
  {\bibfield  {journal} {\bibinfo  {journal} {Phys. Lett.}\ }\textbf {\bibinfo
  {volume} {B349}},\ \bibinfo {pages} {477--483} (\bibinfo {year}
  {1995}{\natexlab{a}})},\ \Eprint {http://arxiv.org/abs/hep-ph/9412257}
  {arXiv:hep-ph/9412257 [hep-ph]} \BibitemShut {NoStop}%
\bibitem [{\citenamefont {Gusynin}\ \emph {et~al.}(1996)\citenamefont
  {Gusynin}, \citenamefont {Miransky},\ and\ \citenamefont
  {Shovkovy}}]{Gusynin:1995nb}%
  \BibitemOpen
  \bibfield  {author} {\bibinfo {author} {\bibfnamefont {V.~P.}\ \bibnamefont
  {Gusynin}}, \bibinfo {author} {\bibfnamefont {V.~A.}\ \bibnamefont
  {Miransky}}, \ and\ \bibinfo {author} {\bibfnamefont {I.~A.}\ \bibnamefont
  {Shovkovy}},\ }\bibfield  {title} {\enquote {\bibinfo {title} {{Dimensional
  reduction and catalysis of dynamical symmetry breaking by a magnetic
  field}},}\ }\href {\doibase 10.1016/0550-3213(96)00021-1} {\bibfield
  {journal} {\bibinfo  {journal} {Nucl. Phys.}\ }\textbf {\bibinfo {volume}
  {B462}},\ \bibinfo {pages} {249--290} (\bibinfo {year} {1996})},\ \Eprint
  {http://arxiv.org/abs/hep-ph/9509320} {arXiv:hep-ph/9509320 [hep-ph]}
  \BibitemShut {NoStop}%
\bibitem [{\citenamefont {Cohen}\ \emph {et~al.}(2007)\citenamefont {Cohen},
  \citenamefont {McGady},\ and\ \citenamefont {Werbos}}]{Cohen:2007bt}%
  \BibitemOpen
  \bibfield  {author} {\bibinfo {author} {\bibfnamefont {Thomas~D.}\
  \bibnamefont {Cohen}}, \bibinfo {author} {\bibfnamefont {David~A.}\
  \bibnamefont {McGady}}, \ and\ \bibinfo {author} {\bibfnamefont
  {Elizabeth~S.}\ \bibnamefont {Werbos}},\ }\bibfield  {title} {\enquote
  {\bibinfo {title} {{The Chiral condensate in a constant electromagnetic
  field}},}\ }\href {\doibase 10.1103/PhysRevC.76.055201} {\bibfield  {journal}
  {\bibinfo  {journal} {Phys. Rev.}\ }\textbf {\bibinfo {volume} {C76}},\
  \bibinfo {pages} {055201} (\bibinfo {year} {2007})},\ \Eprint
  {http://arxiv.org/abs/0706.3208} {arXiv:0706.3208 [hep-ph]} \BibitemShut
  {NoStop}%
\bibitem [{\citenamefont {Mizher}\ \emph {et~al.}(2010)\citenamefont {Mizher},
  \citenamefont {Chernodub},\ and\ \citenamefont {Fraga}}]{Mizher:2010zb}%
  \BibitemOpen
  \bibfield  {author} {\bibinfo {author} {\bibfnamefont {Ana~Julia}\
  \bibnamefont {Mizher}}, \bibinfo {author} {\bibfnamefont {M.~N.}\
  \bibnamefont {Chernodub}}, \ and\ \bibinfo {author} {\bibfnamefont
  {Eduardo~S.}\ \bibnamefont {Fraga}},\ }\bibfield  {title} {\enquote {\bibinfo
  {title} {{Phase diagram of hot QCD in an external magnetic field: possible
  splitting of deconfinement and chiral transitions}},}\ }\href {\doibase
  10.1103/PhysRevD.82.105016} {\bibfield  {journal} {\bibinfo  {journal} {Phys.
  Rev.}\ }\textbf {\bibinfo {volume} {D82}},\ \bibinfo {pages} {105016}
  (\bibinfo {year} {2010})},\ \Eprint {http://arxiv.org/abs/1004.2712}
  {arXiv:1004.2712 [hep-ph]} \BibitemShut {NoStop}%
\bibitem [{\citenamefont {Skokov}(2012)}]{Skokov:2011ib}%
  \BibitemOpen
  \bibfield  {author} {\bibinfo {author} {\bibfnamefont {V.}~\bibnamefont
  {Skokov}},\ }\bibfield  {title} {\enquote {\bibinfo {title} {{Phase diagram
  in an external magnetic field beyond a mean-field approximation}},}\ }\href
  {\doibase 10.1103/PhysRevD.85.034026} {\bibfield  {journal} {\bibinfo
  {journal} {Phys. Rev.}\ }\textbf {\bibinfo {volume} {D85}},\ \bibinfo {pages}
  {034026} (\bibinfo {year} {2012})},\ \Eprint {http://arxiv.org/abs/1112.5137}
  {arXiv:1112.5137 [hep-ph]} \BibitemShut {NoStop}%
\bibitem [{\citenamefont {Fukushima}\ and\ \citenamefont
  {Pawlowski}(2012)}]{Fukushima:2012xw}%
  \BibitemOpen
  \bibfield  {author} {\bibinfo {author} {\bibfnamefont {Kenji}\ \bibnamefont
  {Fukushima}}\ and\ \bibinfo {author} {\bibfnamefont {Jan~M.}\ \bibnamefont
  {Pawlowski}},\ }\bibfield  {title} {\enquote {\bibinfo {title} {{Magnetic
  catalysis in hot and dense quark matter and quantum fluctuations}},}\ }\href
  {\doibase 10.1103/PhysRevD.86.076013} {\bibfield  {journal} {\bibinfo
  {journal} {Phys. Rev.}\ }\textbf {\bibinfo {volume} {D86}},\ \bibinfo {pages}
  {076013} (\bibinfo {year} {2012})},\ \Eprint {http://arxiv.org/abs/1203.4330}
  {arXiv:1203.4330 [hep-ph]} \BibitemShut {NoStop}%
\bibitem [{\citenamefont {Andersen}\ \emph {et~al.}(2016)\citenamefont
  {Andersen}, \citenamefont {Naylor},\ and\ \citenamefont
  {Tranberg}}]{Andersen:2014xxa}%
  \BibitemOpen
  \bibfield  {author} {\bibinfo {author} {\bibfnamefont {Jens~O.}\ \bibnamefont
  {Andersen}}, \bibinfo {author} {\bibfnamefont {William~R.}\ \bibnamefont
  {Naylor}}, \ and\ \bibinfo {author} {\bibfnamefont {Anders}\ \bibnamefont
  {Tranberg}},\ }\bibfield  {title} {\enquote {\bibinfo {title} {{Phase diagram
  of QCD in a magnetic field: A review}},}\ }\href {\doibase
  10.1103/RevModPhys.88.025001} {\bibfield  {journal} {\bibinfo  {journal}
  {Rev. Mod. Phys.}\ }\textbf {\bibinfo {volume} {88}},\ \bibinfo {pages}
  {025001} (\bibinfo {year} {2016})},\ \Eprint {http://arxiv.org/abs/1411.7176}
  {arXiv:1411.7176 [hep-ph]} \BibitemShut {NoStop}%
\bibitem [{\citenamefont {Gusynin}\ \emph
  {et~al.}(1995{\natexlab{b}})\citenamefont {Gusynin}, \citenamefont
  {Miransky},\ and\ \citenamefont {Shovkovy}}]{Gusynin:1995gt}%
  \BibitemOpen
  \bibfield  {author} {\bibinfo {author} {\bibfnamefont {V.~P.}\ \bibnamefont
  {Gusynin}}, \bibinfo {author} {\bibfnamefont {V.~A.}\ \bibnamefont
  {Miransky}}, \ and\ \bibinfo {author} {\bibfnamefont {I.~A.}\ \bibnamefont
  {Shovkovy}},\ }\bibfield  {title} {\enquote {\bibinfo {title} {{Dynamical
  chiral symmetry breaking by a magnetic field in QED}},}\ }\href {\doibase
  10.1103/PhysRevD.52.4747} {\bibfield  {journal} {\bibinfo  {journal} {Phys.
  Rev.}\ }\textbf {\bibinfo {volume} {D52}},\ \bibinfo {pages} {4747--4751}
  (\bibinfo {year} {1995}{\natexlab{b}})},\ \Eprint
  {http://arxiv.org/abs/hep-ph/9501304} {arXiv:hep-ph/9501304 [hep-ph]}
  \BibitemShut {NoStop}%
\bibitem [{\citenamefont {Gross}\ \emph {et~al.}(1981)\citenamefont {Gross},
  \citenamefont {Pisarski},\ and\ \citenamefont {Yaffe}}]{Gross:1980br}%
  \BibitemOpen
  \bibfield  {author} {\bibinfo {author} {\bibfnamefont {David~J.}\
  \bibnamefont {Gross}}, \bibinfo {author} {\bibfnamefont {Robert~D.}\
  \bibnamefont {Pisarski}}, \ and\ \bibinfo {author} {\bibfnamefont
  {Laurence~G.}\ \bibnamefont {Yaffe}},\ }\bibfield  {title} {\enquote
  {\bibinfo {title} {{QCD and Instantons at Finite Temperature}},}\ }\href
  {\doibase 10.1103/RevModPhys.53.43} {\bibfield  {journal} {\bibinfo
  {journal} {Rev. Mod. Phys.}\ }\textbf {\bibinfo {volume} {53}},\ \bibinfo
  {pages} {43} (\bibinfo {year} {1981})}\BibitemShut {NoStop}%
\bibitem [{\citenamefont {Weiss}(1981)}]{Weiss:1980rj}%
  \BibitemOpen
  \bibfield  {author} {\bibinfo {author} {\bibfnamefont {Nathan}\ \bibnamefont
  {Weiss}},\ }\bibfield  {title} {\enquote {\bibinfo {title} {{The Effective
  Potential for the Order Parameter of Gauge Theories at Finite
  Temperature}},}\ }\href {\doibase 10.1103/PhysRevD.24.475} {\bibfield
  {journal} {\bibinfo  {journal} {Phys. Rev.}\ }\textbf {\bibinfo {volume}
  {D24}},\ \bibinfo {pages} {475} (\bibinfo {year} {1981})}\BibitemShut
  {NoStop}%
\bibitem [{\citenamefont {Gies}(2001)}]{Gies:2000dw}%
  \BibitemOpen
  \bibfield  {author} {\bibinfo {author} {\bibfnamefont {Holger}\ \bibnamefont
  {Gies}},\ }\bibfield  {title} {\enquote {\bibinfo {title} {{Effective action
  for the order parameter of the deconfinement transition of Yang-Mills
  theories}},}\ }\href {\doibase 10.1103/PhysRevD.63.025013} {\bibfield
  {journal} {\bibinfo  {journal} {Phys. Rev.}\ }\textbf {\bibinfo {volume}
  {D63}},\ \bibinfo {pages} {025013} (\bibinfo {year} {2001})},\ \Eprint
  {http://arxiv.org/abs/hep-th/0005252} {arXiv:hep-th/0005252 [hep-th]}
  \BibitemShut {NoStop}%
\bibitem [{\citenamefont {Ozaki}\ \emph {et~al.}(2015)\citenamefont {Ozaki},
  \citenamefont {Arai}, \citenamefont {Hattori},\ and\ \citenamefont
  {Itakura}}]{Ozaki:2015yja}%
  \BibitemOpen
  \bibfield  {author} {\bibinfo {author} {\bibfnamefont {Sho}\ \bibnamefont
  {Ozaki}}, \bibinfo {author} {\bibfnamefont {Takashi}\ \bibnamefont {Arai}},
  \bibinfo {author} {\bibfnamefont {Koichi}\ \bibnamefont {Hattori}}, \ and\
  \bibinfo {author} {\bibfnamefont {Kazunori}\ \bibnamefont {Itakura}},\
  }\bibfield  {title} {\enquote {\bibinfo {title} {{Euler-Heisenberg-Weiss
  action for QCD+QED}},}\ }\href {\doibase 10.1103/PhysRevD.92.016002}
  {\bibfield  {journal} {\bibinfo  {journal} {Phys. Rev.}\ }\textbf {\bibinfo
  {volume} {D92}},\ \bibinfo {pages} {016002} (\bibinfo {year} {2015})},\
  \Eprint {http://arxiv.org/abs/1504.07532} {arXiv:1504.07532 [hep-ph]}
  \BibitemShut {NoStop}%
\bibitem [{\citenamefont {D'Elia}\ \emph {et~al.}(2018)\citenamefont {D'Elia},
  \citenamefont {Manigrasso}, \citenamefont {Negro},\ and\ \citenamefont
  {Sanfilippo}}]{DElia:2018xwo}%
  \BibitemOpen
  \bibfield  {author} {\bibinfo {author} {\bibfnamefont {Massimo}\ \bibnamefont
  {D'Elia}}, \bibinfo {author} {\bibfnamefont {Floriano}\ \bibnamefont
  {Manigrasso}}, \bibinfo {author} {\bibfnamefont {Francesco}\ \bibnamefont
  {Negro}}, \ and\ \bibinfo {author} {\bibfnamefont {Francesco}\ \bibnamefont
  {Sanfilippo}},\ }\bibfield  {title} {\enquote {\bibinfo {title} {{QCD phase
  diagram in a magnetic background for different values of the pion mass}},}\
  }\href {\doibase 10.1103/PhysRevD.98.054509} {\bibfield  {journal} {\bibinfo
  {journal} {Phys. Rev.}\ }\textbf {\bibinfo {volume} {D98}},\ \bibinfo {pages}
  {054509} (\bibinfo {year} {2018})},\ \Eprint
  {http://arxiv.org/abs/1808.07008} {arXiv:1808.07008 [hep-lat]} \BibitemShut
  {NoStop}%
\bibitem [{\citenamefont {Tomiya}\ \emph {et~al.}(2019)\citenamefont {Tomiya},
  \citenamefont {Ding}, \citenamefont {Wang}, \citenamefont {Zhang},
  \citenamefont {Mukherjee},\ and\ \citenamefont {Schmidt}}]{Tomiya:2019nym}%
  \BibitemOpen
  \bibfield  {author} {\bibinfo {author} {\bibfnamefont {Akio}\ \bibnamefont
  {Tomiya}}, \bibinfo {author} {\bibfnamefont {Heng-Tong}\ \bibnamefont
  {Ding}}, \bibinfo {author} {\bibfnamefont {Xiao-Dan}\ \bibnamefont {Wang}},
  \bibinfo {author} {\bibfnamefont {Yu}~\bibnamefont {Zhang}}, \bibinfo
  {author} {\bibfnamefont {Swagato}\ \bibnamefont {Mukherjee}}, \ and\ \bibinfo
  {author} {\bibfnamefont {Christian}\ \bibnamefont {Schmidt}},\ }\bibfield
  {title} {\enquote {\bibinfo {title} {{Phase structure of three flavor QCD in
  external magnetic fields using HISQ fermions}},}\ }\href@noop {} {\bibfield
  {journal} {\bibinfo  {journal} {Submitted to: PoS}\ } (\bibinfo {year}
  {2019})},\ \Eprint {http://arxiv.org/abs/1904.01276} {arXiv:1904.01276
  [hep-lat]} \BibitemShut {NoStop}%
\bibitem [{\citenamefont {Endrodi}\ \emph {et~al.}(2019)\citenamefont
  {Endrodi}, \citenamefont {Giordano}, \citenamefont {Katz}, \citenamefont
  {Kovacs},\ and\ \citenamefont {Pittler}}]{Endrodi:2019zrl}%
  \BibitemOpen
  \bibfield  {author} {\bibinfo {author} {\bibfnamefont {Gergely}\ \bibnamefont
  {Endrodi}}, \bibinfo {author} {\bibfnamefont {Matteo}\ \bibnamefont
  {Giordano}}, \bibinfo {author} {\bibfnamefont {Sandor~D.}\ \bibnamefont
  {Katz}}, \bibinfo {author} {\bibfnamefont {Tamas~G.}\ \bibnamefont {Kovacs}},
  \ and\ \bibinfo {author} {\bibfnamefont {Ferenc}\ \bibnamefont {Pittler}},\
  }\bibfield  {title} {\enquote {\bibinfo {title} {{Magnetic catalysis and
  inverse catalysis for heavy pions}},}\ }\href@noop {} {\  (\bibinfo {year}
  {2019})},\ \Eprint {http://arxiv.org/abs/1904.10296} {arXiv:1904.10296
  [hep-lat]} \BibitemShut {NoStop}%
\bibitem [{\citenamefont {Hattori}\ and\ \citenamefont
  {Itakura}(2013{\natexlab{a}})}]{Hattori:2012je}%
  \BibitemOpen
  \bibfield  {author} {\bibinfo {author} {\bibfnamefont {Koichi}\ \bibnamefont
  {Hattori}}\ and\ \bibinfo {author} {\bibfnamefont {Kazunori}\ \bibnamefont
  {Itakura}},\ }\bibfield  {title} {\enquote {\bibinfo {title} {{Vacuum
  birefringence in strong magnetic fields: (I) Photon polarization tensor with
  all the Landau levels}},}\ }\href {\doibase 10.1016/j.aop.2012.11.010}
  {\bibfield  {journal} {\bibinfo  {journal} {Annals Phys.}\ }\textbf {\bibinfo
  {volume} {330}},\ \bibinfo {pages} {23--54} (\bibinfo {year}
  {2013}{\natexlab{a}})},\ \Eprint {http://arxiv.org/abs/1209.2663}
  {arXiv:1209.2663 [hep-ph]} \BibitemShut {NoStop}%
\bibitem [{\citenamefont {Hattori}\ and\ \citenamefont
  {Itakura}(2013{\natexlab{b}})}]{Hattori:2012ny}%
  \BibitemOpen
  \bibfield  {author} {\bibinfo {author} {\bibfnamefont {Koichi}\ \bibnamefont
  {Hattori}}\ and\ \bibinfo {author} {\bibfnamefont {Kazunori}\ \bibnamefont
  {Itakura}},\ }\bibfield  {title} {\enquote {\bibinfo {title} {{Vacuum
  birefringence in strong magnetic fields: (II) Complex refractive index from
  the lowest Landau level}},}\ }\href {\doibase 10.1016/j.aop.2013.03.016}
  {\bibfield  {journal} {\bibinfo  {journal} {Annals Phys.}\ }\textbf {\bibinfo
  {volume} {334}},\ \bibinfo {pages} {58--82} (\bibinfo {year}
  {2013}{\natexlab{b}})},\ \Eprint {http://arxiv.org/abs/1212.1897}
  {arXiv:1212.1897 [hep-ph]} \BibitemShut {NoStop}%
\bibitem [{\citenamefont {Ishikawa}\ \emph {et~al.}(2013)\citenamefont
  {Ishikawa}, \citenamefont {Kimura}, \citenamefont {Shigaki},\ and\
  \citenamefont {Tsuji}}]{Ishikawa:2013fxa}%
  \BibitemOpen
  \bibfield  {author} {\bibinfo {author} {\bibfnamefont {Ken-Ichi}\
  \bibnamefont {Ishikawa}}, \bibinfo {author} {\bibfnamefont {Daiji}\
  \bibnamefont {Kimura}}, \bibinfo {author} {\bibfnamefont {Kenta}\
  \bibnamefont {Shigaki}}, \ and\ \bibinfo {author} {\bibfnamefont {Asako}\
  \bibnamefont {Tsuji}},\ }\bibfield  {title} {\enquote {\bibinfo {title} {{A
  numerical evaluation of vacuum polarization tensor in constant external
  magnetic fields}},}\ }\href {\doibase 10.1142/S0217751X13501005} {\bibfield
  {journal} {\bibinfo  {journal} {Int. J. Mod. Phys.}\ }\textbf {\bibinfo
  {volume} {A28}},\ \bibinfo {pages} {1350100} (\bibinfo {year} {2013})},\
  \Eprint {http://arxiv.org/abs/1304.3655} {arXiv:1304.3655 [hep-ph]}
  \BibitemShut {NoStop}%
\bibitem [{\citenamefont {Sheng}\ \emph {et~al.}(2019)\citenamefont {Sheng},
  \citenamefont {Fang}, \citenamefont {Wang},\ and\ \citenamefont
  {Rischke}}]{Sheng:2018jwf}%
  \BibitemOpen
  \bibfield  {author} {\bibinfo {author} {\bibfnamefont {Xin-Li}\ \bibnamefont
  {Sheng}}, \bibinfo {author} {\bibfnamefont {Ren-Hong}\ \bibnamefont {Fang}},
  \bibinfo {author} {\bibfnamefont {Qun}\ \bibnamefont {Wang}}, \ and\ \bibinfo
  {author} {\bibfnamefont {Dirk~H.}\ \bibnamefont {Rischke}},\ }\bibfield
  {title} {\enquote {\bibinfo {title} {{Wigner function and pair production in
  parallel electric and magnetic fields}},}\ }\href {\doibase
  10.1103/PhysRevD.99.056004} {\bibfield  {journal} {\bibinfo  {journal} {Phys.
  Rev.}\ }\textbf {\bibinfo {volume} {D99}},\ \bibinfo {pages} {056004}
  (\bibinfo {year} {2019})},\ \Eprint {http://arxiv.org/abs/1812.01146}
  {arXiv:1812.01146 [hep-ph]} \BibitemShut {NoStop}%
\bibitem [{\citenamefont {Brown}\ and\ \citenamefont
  {Duff}(1975)}]{Brown:1975bc}%
  \BibitemOpen
  \bibfield  {author} {\bibinfo {author} {\bibfnamefont {M.~R.}\ \bibnamefont
  {Brown}}\ and\ \bibinfo {author} {\bibfnamefont {M.~J.}\ \bibnamefont
  {Duff}},\ }\bibfield  {title} {\enquote {\bibinfo {title} {{Exact Results for
  Effective Lagrangians}},}\ }\href {\doibase 10.1103/PhysRevD.11.2124}
  {\bibfield  {journal} {\bibinfo  {journal} {Phys. Rev.}\ }\textbf {\bibinfo
  {volume} {D11}},\ \bibinfo {pages} {2124--2135} (\bibinfo {year}
  {1975})}\BibitemShut {NoStop}%
\bibitem [{\citenamefont {Duff}\ and\ \citenamefont
  {Ramon-Medrano}(1975)}]{Duff:1975ue}%
  \BibitemOpen
  \bibfield  {author} {\bibinfo {author} {\bibfnamefont {M.~J.}\ \bibnamefont
  {Duff}}\ and\ \bibinfo {author} {\bibfnamefont {M.}~\bibnamefont
  {Ramon-Medrano}},\ }\bibfield  {title} {\enquote {\bibinfo {title} {{On the
  Effective Lagrangian for the Yang-Mills Field}},}\ }\href {\doibase
  10.1103/PhysRevD.12.3357} {\bibfield  {journal} {\bibinfo  {journal} {Phys.
  Rev.}\ }\textbf {\bibinfo {volume} {D12}},\ \bibinfo {pages} {3357} (\bibinfo
  {year} {1975})}\BibitemShut {NoStop}%
\bibitem [{\citenamefont {Dittrich}\ and\ \citenamefont
  {Reuter}(1985)}]{Dittrich:1985yb}%
  \BibitemOpen
  \bibfield  {author} {\bibinfo {author} {\bibfnamefont {W.}~\bibnamefont
  {Dittrich}}\ and\ \bibinfo {author} {\bibfnamefont {M.}~\bibnamefont
  {Reuter}},\ }\bibfield  {title} {\enquote {\bibinfo {title} {{EFFECTIVE
  LAGRANGIANS IN QUANTUM ELECTRODYNAMICS}},}\ }\href@noop {} {\bibfield
  {journal} {\bibinfo  {journal} {Lect. Notes Phys.}\ }\textbf {\bibinfo
  {volume} {220}},\ \bibinfo {pages} {1--244} (\bibinfo {year}
  {1985})}\BibitemShut {NoStop}%
\bibitem [{\citenamefont {Nikishov}(1969)}]{Nikishov:1969tt}%
  \BibitemOpen
  \bibfield  {author} {\bibinfo {author} {\bibfnamefont {A.~I.}\ \bibnamefont
  {Nikishov}},\ }\bibfield  {title} {\enquote {\bibinfo {title} {{Pair
  production by a constant external field}},}\ }\href@noop {} {\bibfield
  {journal} {\bibinfo  {journal} {Zh. Eksp. Teor. Fiz.}\ }\textbf {\bibinfo
  {volume} {57}},\ \bibinfo {pages} {1210--1216} (\bibinfo {year}
  {1969})}\BibitemShut {NoStop}%
\bibitem [{\citenamefont {Holstein}(1999)}]{Holstein:1999ta}%
  \BibitemOpen
  \bibfield  {author} {\bibinfo {author} {\bibfnamefont {Barry~R.}\
  \bibnamefont {Holstein}},\ }\bibfield  {title} {\enquote {\bibinfo {title}
  {{Strong field pair production}},}\ }\href {\doibase 10.1119/1.19313}
  {\bibfield  {journal} {\bibinfo  {journal} {Am. J. Phys.}\ }\textbf {\bibinfo
  {volume} {67}},\ \bibinfo {pages} {499--507} (\bibinfo {year}
  {1999})}\BibitemShut {NoStop}%
\bibitem [{\citenamefont {Cohen}\ and\ \citenamefont
  {McGady}(2008)}]{Cohen:2008wz}%
  \BibitemOpen
  \bibfield  {author} {\bibinfo {author} {\bibfnamefont {Thomas~D.}\
  \bibnamefont {Cohen}}\ and\ \bibinfo {author} {\bibfnamefont {David~A.}\
  \bibnamefont {McGady}},\ }\bibfield  {title} {\enquote {\bibinfo {title}
  {{The Schwinger mechanism revisited}},}\ }\href {\doibase
  10.1103/PhysRevD.78.036008} {\bibfield  {journal} {\bibinfo  {journal} {Phys.
  Rev.}\ }\textbf {\bibinfo {volume} {D78}},\ \bibinfo {pages} {036008}
  (\bibinfo {year} {2008})},\ \Eprint {http://arxiv.org/abs/0807.1117}
  {arXiv:0807.1117 [hep-ph]} \BibitemShut {NoStop}%
\bibitem [{\citenamefont {Tanji}(2009)}]{Tanji:2008ku}%
  \BibitemOpen
  \bibfield  {author} {\bibinfo {author} {\bibfnamefont {Naoto}\ \bibnamefont
  {Tanji}},\ }\bibfield  {title} {\enquote {\bibinfo {title} {{Dynamical view
  of pair creation in uniform electric and magnetic fields}},}\ }\href
  {\doibase 10.1016/j.aop.2009.03.012} {\bibfield  {journal} {\bibinfo
  {journal} {Annals Phys.}\ }\textbf {\bibinfo {volume} {324}},\ \bibinfo
  {pages} {1691--1736} (\bibinfo {year} {2009})},\ \Eprint
  {http://arxiv.org/abs/0810.4429} {arXiv:0810.4429 [hep-ph]} \BibitemShut
  {NoStop}%
\bibitem [{\citenamefont {Hattori}\ \emph {et~al.}(2017)\citenamefont
  {Hattori}, \citenamefont {Itakura},\ and\ \citenamefont
  {Ozaki}}]{Hattori:2017qio}%
  \BibitemOpen
  \bibfield  {author} {\bibinfo {author} {\bibfnamefont {Koichi}\ \bibnamefont
  {Hattori}}, \bibinfo {author} {\bibfnamefont {Kazunori}\ \bibnamefont
  {Itakura}}, \ and\ \bibinfo {author} {\bibfnamefont {Sho}\ \bibnamefont
  {Ozaki}},\ }\bibfield  {title} {\enquote {\bibinfo {title} {{Anatomy of the
  magnetic catalysis by renormalization-group method}},}\ }\href {\doibase
  10.1016/j.physletb.2017.11.004} {\bibfield  {journal} {\bibinfo  {journal}
  {Phys. Lett.}\ }\textbf {\bibinfo {volume} {B775}},\ \bibinfo {pages}
  {283--289} (\bibinfo {year} {2017})},\ \Eprint
  {http://arxiv.org/abs/1706.04913} {arXiv:1706.04913 [hep-ph]} \BibitemShut
  {NoStop}%
\bibitem [{\citenamefont {Ambjorn}\ and\ \citenamefont
  {Hughes}(1983)}]{Ambjorn:1982bp}%
  \BibitemOpen
  \bibfield  {author} {\bibinfo {author} {\bibfnamefont {Jan}\ \bibnamefont
  {Ambjorn}}\ and\ \bibinfo {author} {\bibfnamefont {Richard~J.}\ \bibnamefont
  {Hughes}},\ }\bibfield  {title} {\enquote {\bibinfo {title} {{Canonical
  Quantization in Nonabelian Background Fields. 1.}}}\ }\href {\doibase
  10.1016/0003-4916(83)90187-2} {\bibfield  {journal} {\bibinfo  {journal}
  {Annals Phys.}\ }\textbf {\bibinfo {volume} {145}},\ \bibinfo {pages} {340}
  (\bibinfo {year} {1983})}\BibitemShut {NoStop}%
\bibitem [{\citenamefont {Suganuma}\ and\ \citenamefont
  {Tatsumi}(1993)}]{Suganuma:1991ha}%
  \BibitemOpen
  \bibfield  {author} {\bibinfo {author} {\bibfnamefont {Hideo}\ \bibnamefont
  {Suganuma}}\ and\ \bibinfo {author} {\bibfnamefont {Toshitaka}\ \bibnamefont
  {Tatsumi}},\ }\bibfield  {title} {\enquote {\bibinfo {title} {{Chiral
  symmetry and quark - anti-quark pair creation in a strong color
  electromagnetic field}},}\ }\href {\doibase 10.1143/PTP.90.379} {\bibfield
  {journal} {\bibinfo  {journal} {Prog. Theor. Phys.}\ }\textbf {\bibinfo
  {volume} {90}},\ \bibinfo {pages} {379--404} (\bibinfo {year}
  {1993})}\BibitemShut {NoStop}%
\bibitem [{\citenamefont {Nayak}\ and\ \citenamefont {van
  Nieuwenhuizen}(2005)}]{Nayak:2005yv}%
  \BibitemOpen
  \bibfield  {author} {\bibinfo {author} {\bibfnamefont {Gouranga~C.}\
  \bibnamefont {Nayak}}\ and\ \bibinfo {author} {\bibfnamefont {Peter}\
  \bibnamefont {van Nieuwenhuizen}},\ }\bibfield  {title} {\enquote {\bibinfo
  {title} {{Soft-gluon production due to a gluon loop in a constant
  chromo-electric background field}},}\ }\href {\doibase
  10.1103/PhysRevD.71.125001} {\bibfield  {journal} {\bibinfo  {journal} {Phys.
  Rev.}\ }\textbf {\bibinfo {volume} {D71}},\ \bibinfo {pages} {125001}
  (\bibinfo {year} {2005})},\ \Eprint {http://arxiv.org/abs/hep-ph/0504070}
  {arXiv:hep-ph/0504070 [hep-ph]} \BibitemShut {NoStop}%
\bibitem [{\citenamefont {Nayak}(2005)}]{Nayak:2005pf}%
  \BibitemOpen
  \bibfield  {author} {\bibinfo {author} {\bibfnamefont {Gouranga~C.}\
  \bibnamefont {Nayak}},\ }\bibfield  {title} {\enquote {\bibinfo {title}
  {{Non-perturbative quark-antiquark production from a constant chromo-electric
  field via the Schwinger mechanism}},}\ }\href {\doibase
  10.1103/PhysRevD.72.125010} {\bibfield  {journal} {\bibinfo  {journal} {Phys.
  Rev.}\ }\textbf {\bibinfo {volume} {D72}},\ \bibinfo {pages} {125010}
  (\bibinfo {year} {2005})},\ \Eprint {http://arxiv.org/abs/hep-ph/0510052}
  {arXiv:hep-ph/0510052 [hep-ph]} \BibitemShut {NoStop}%
\bibitem [{\citenamefont {Tanji}(2010)}]{Tanji:2010eu}%
  \BibitemOpen
  \bibfield  {author} {\bibinfo {author} {\bibfnamefont {Noato}\ \bibnamefont
  {Tanji}},\ }\bibfield  {title} {\enquote {\bibinfo {title} {{Quark pair
  creation in color electric fields and effects of magnetic fields}},}\ }\href
  {\doibase 10.1016/j.aop.2010.03.012} {\bibfield  {journal} {\bibinfo
  {journal} {Annals Phys.}\ }\textbf {\bibinfo {volume} {325}},\ \bibinfo
  {pages} {2018--2040} (\bibinfo {year} {2010})},\ \Eprint
  {http://arxiv.org/abs/1002.3143} {arXiv:1002.3143 [hep-ph]} \BibitemShut
  {NoStop}%
\bibitem [{\citenamefont {Tanji}\ and\ \citenamefont
  {Itakura}(2012)}]{Tanji:2011di}%
  \BibitemOpen
  \bibfield  {author} {\bibinfo {author} {\bibfnamefont {Naoto}\ \bibnamefont
  {Tanji}}\ and\ \bibinfo {author} {\bibfnamefont {Kazunori}\ \bibnamefont
  {Itakura}},\ }\bibfield  {title} {\enquote {\bibinfo {title} {{Schwinger
  mechanism enhanced by the Nielsen-Olesen instability}},}\ }\href {\doibase
  10.1016/j.physletb.2012.05.043} {\bibfield  {journal} {\bibinfo  {journal}
  {Phys. Lett.}\ }\textbf {\bibinfo {volume} {B713}},\ \bibinfo {pages}
  {117--121} (\bibinfo {year} {2012})},\ \Eprint
  {http://arxiv.org/abs/1111.6772} {arXiv:1111.6772 [hep-ph]} \BibitemShut
  {NoStop}%
\bibitem [{\citenamefont {Claudson}\ \emph {et~al.}(1980)\citenamefont
  {Claudson}, \citenamefont {Yildiz},\ and\ \citenamefont
  {Cox}}]{Claudson:1980yz}%
  \BibitemOpen
  \bibfield  {author} {\bibinfo {author} {\bibfnamefont {M.}~\bibnamefont
  {Claudson}}, \bibinfo {author} {\bibfnamefont {A.}~\bibnamefont {Yildiz}}, \
  and\ \bibinfo {author} {\bibfnamefont {Paul~H.}\ \bibnamefont {Cox}},\
  }\bibfield  {title} {\enquote {\bibinfo {title} {{VACUUM BEHAVIOR IN QUANTUM
  CHROMODYNAMICS. II}},}\ }\href {\doibase 10.1103/PhysRevD.22.2022} {\bibfield
   {journal} {\bibinfo  {journal} {Phys. Rev.}\ }\textbf {\bibinfo {volume}
  {D22}},\ \bibinfo {pages} {2022--2026} (\bibinfo {year} {1980})}\BibitemShut
  {NoStop}%
\bibitem [{\citenamefont {Leutwyler}(1981)}]{Leutwyler:1980ma}%
  \BibitemOpen
  \bibfield  {author} {\bibinfo {author} {\bibfnamefont {H.}~\bibnamefont
  {Leutwyler}},\ }\bibfield  {title} {\enquote {\bibinfo {title} {{Constant
  Gauge Fields and their Quantum Fluctuations}},}\ }\href {\doibase
  10.1016/0550-3213(81)90252-2} {\bibfield  {journal} {\bibinfo  {journal}
  {Nucl. Phys.}\ }\textbf {\bibinfo {volume} {B179}},\ \bibinfo {pages}
  {129--170} (\bibinfo {year} {1981})}\BibitemShut {NoStop}%
\bibitem [{\citenamefont {Schanbacher}(1982)}]{Schanbacher:1980vq}%
  \BibitemOpen
  \bibfield  {author} {\bibinfo {author} {\bibfnamefont {Volker}\ \bibnamefont
  {Schanbacher}},\ }\bibfield  {title} {\enquote {\bibinfo {title} {{Gluon
  Propagator and Effective Lagrangian in {QCD}}},}\ }\href {\doibase
  10.1103/PhysRevD.26.489} {\bibfield  {journal} {\bibinfo  {journal} {Phys.
  Rev.}\ }\textbf {\bibinfo {volume} {D26}},\ \bibinfo {pages} {489} (\bibinfo
  {year} {1982})}\BibitemShut {NoStop}%
\bibitem [{\citenamefont {Ambjorn}\ and\ \citenamefont
  {Hughes}(1982)}]{Ambjorn:1982nd}%
  \BibitemOpen
  \bibfield  {author} {\bibinfo {author} {\bibfnamefont {Jan}\ \bibnamefont
  {Ambjorn}}\ and\ \bibinfo {author} {\bibfnamefont {Richard~J.}\ \bibnamefont
  {Hughes}},\ }\bibfield  {title} {\enquote {\bibinfo {title} {{Particle
  Creation in Color Electric Fields}},}\ }\href {\doibase
  10.1016/0370-2693(82)90045-4} {\bibfield  {journal} {\bibinfo  {journal}
  {Phys. Lett.}\ }\textbf {\bibinfo {volume} {113B}},\ \bibinfo {pages}
  {305--310} (\bibinfo {year} {1982})}\BibitemShut {NoStop}%
\bibitem [{\citenamefont {Ambjorn}\ \emph {et~al.}(1983)\citenamefont
  {Ambjorn}, \citenamefont {Hughes},\ and\ \citenamefont
  {Nielsen}}]{Ambjorn:1983ne}%
  \BibitemOpen
  \bibfield  {author} {\bibinfo {author} {\bibfnamefont {Jan}\ \bibnamefont
  {Ambjorn}}, \bibinfo {author} {\bibfnamefont {R.~J.}\ \bibnamefont {Hughes}},
  \ and\ \bibinfo {author} {\bibfnamefont {N.~K.}\ \bibnamefont {Nielsen}},\
  }\bibfield  {title} {\enquote {\bibinfo {title} {{Action Principle of
  Bogolyubov Coefficients}},}\ }\href {\doibase 10.1016/0003-4916(83)90005-2}
  {\bibfield  {journal} {\bibinfo  {journal} {Annals Phys.}\ }\textbf {\bibinfo
  {volume} {150}},\ \bibinfo {pages} {92} (\bibinfo {year} {1983})}\BibitemShut
  {NoStop}%
\bibitem [{\citenamefont {Elizalde}\ and\ \citenamefont
  {Soto}(1985)}]{Elizalde:1984zv}%
  \BibitemOpen
  \bibfield  {author} {\bibinfo {author} {\bibfnamefont {E.}~\bibnamefont
  {Elizalde}}\ and\ \bibinfo {author} {\bibfnamefont {J.}~\bibnamefont
  {Soto}},\ }\bibfield  {title} {\enquote {\bibinfo {title} {{Zeta Regularized
  Lagrangians for Massive Quarks in Constant Background Mean Fields}},}\ }\href
  {\doibase 10.1016/0003-4916(85)90233-7} {\bibfield  {journal} {\bibinfo
  {journal} {Annals Phys.}\ }\textbf {\bibinfo {volume} {162}},\ \bibinfo
  {pages} {192} (\bibinfo {year} {1985})}\BibitemShut {NoStop}%
\bibitem [{\citenamefont {Copinger}\ \emph {et~al.}(2018)\citenamefont
  {Copinger}, \citenamefont {Fukushima},\ and\ \citenamefont
  {Pu}}]{Copinger:2018ftr}%
  \BibitemOpen
  \bibfield  {author} {\bibinfo {author} {\bibfnamefont {Patrick}\ \bibnamefont
  {Copinger}}, \bibinfo {author} {\bibfnamefont {Kenji}\ \bibnamefont
  {Fukushima}}, \ and\ \bibinfo {author} {\bibfnamefont {Shi}\ \bibnamefont
  {Pu}},\ }\bibfield  {title} {\enquote {\bibinfo {title} {{Axial Ward identity
  and the Schwinger mechanism -- Applications to the real-time chiral magnetic
  effect and condensates}},}\ }\href {\doibase 10.1103/PhysRevLett.121.261602}
  {\bibfield  {journal} {\bibinfo  {journal} {Phys. Rev. Lett.}\ }\textbf
  {\bibinfo {volume} {121}},\ \bibinfo {pages} {261602} (\bibinfo {year}
  {2018})},\ \Eprint {http://arxiv.org/abs/1807.04416} {arXiv:1807.04416
  [hep-th]} \BibitemShut {NoStop}%
\bibitem [{\citenamefont {Dittrich}(1979)}]{Dittrich:1979ux}%
  \BibitemOpen
  \bibfield  {author} {\bibinfo {author} {\bibfnamefont {Walter}\ \bibnamefont
  {Dittrich}},\ }\bibfield  {title} {\enquote {\bibinfo {title} {{Effective
  Lagrangian at Finite Tempearture}},}\ }\href {\doibase
  10.1103/PhysRevD.19.2385} {\bibfield  {journal} {\bibinfo  {journal} {Phys.
  Rev.}\ }\textbf {\bibinfo {volume} {D19}},\ \bibinfo {pages} {2385} (\bibinfo
  {year} {1979})}\BibitemShut {NoStop}%
\bibitem [{\citenamefont {Muller}\ and\ \citenamefont
  {Rafelski}(1981)}]{Muller:1980kf}%
  \BibitemOpen
  \bibfield  {author} {\bibinfo {author} {\bibfnamefont {Berndt}\ \bibnamefont
  {Muller}}\ and\ \bibinfo {author} {\bibfnamefont {Johann}\ \bibnamefont
  {Rafelski}},\ }\bibfield  {title} {\enquote {\bibinfo {title} {{Temperature
  Dependence of the Bag Constant and the Effective Lagrangian for Gauge Fields
  at Finite Temperatures}},}\ }\href {\doibase 10.1016/0370-2693(81)90502-5}
  {\bibfield  {journal} {\bibinfo  {journal} {Phys. Lett.}\ }\textbf {\bibinfo
  {volume} {101B}},\ \bibinfo {pages} {111--118} (\bibinfo {year}
  {1981})}\BibitemShut {NoStop}%
\bibitem [{\citenamefont {Dittrich}\ and\ \citenamefont
  {Schanbacher}(1981)}]{Dittrich:1980nh}%
  \BibitemOpen
  \bibfield  {author} {\bibinfo {author} {\bibfnamefont {Walter}\ \bibnamefont
  {Dittrich}}\ and\ \bibinfo {author} {\bibfnamefont {Volker}\ \bibnamefont
  {Schanbacher}},\ }\bibfield  {title} {\enquote {\bibinfo {title} {{EFFECTIVE
  QCD LAGRANGIAN AT FINITE TEMPERATURE}},}\ }\href {\doibase
  10.1016/0370-2693(81)90149-0} {\bibfield  {journal} {\bibinfo  {journal}
  {Phys. Lett.}\ }\textbf {\bibinfo {volume} {B100}},\ \bibinfo {pages}
  {415--419} (\bibinfo {year} {1981})}\BibitemShut {NoStop}%
\bibitem [{\citenamefont {Gies}(1999)}]{Gies:1998vt}%
  \BibitemOpen
  \bibfield  {author} {\bibinfo {author} {\bibfnamefont {Holger}\ \bibnamefont
  {Gies}},\ }\bibfield  {title} {\enquote {\bibinfo {title} {{QED effective
  action at finite temperature}},}\ }\href {\doibase
  10.1103/PhysRevD.60.105002} {\bibfield  {journal} {\bibinfo  {journal} {Phys.
  Rev.}\ }\textbf {\bibinfo {volume} {D60}},\ \bibinfo {pages} {105002}
  (\bibinfo {year} {1999})},\ \Eprint {http://arxiv.org/abs/hep-ph/9812436}
  {arXiv:hep-ph/9812436 [hep-ph]} \BibitemShut {NoStop}%
\bibitem [{\citenamefont {Gies}(2000)}]{Gies:1999vb}%
  \BibitemOpen
  \bibfield  {author} {\bibinfo {author} {\bibfnamefont {Holger}\ \bibnamefont
  {Gies}},\ }\bibfield  {title} {\enquote {\bibinfo {title} {{QED effective
  action at finite temperature: Two loop dominance}},}\ }\href {\doibase
  10.1103/PhysRevD.61.085021} {\bibfield  {journal} {\bibinfo  {journal} {Phys.
  Rev.}\ }\textbf {\bibinfo {volume} {D61}},\ \bibinfo {pages} {085021}
  (\bibinfo {year} {2000})},\ \Eprint {http://arxiv.org/abs/hep-ph/9909500}
  {arXiv:hep-ph/9909500 [hep-ph]} \BibitemShut {NoStop}%
\bibitem [{\citenamefont {Ritus}(1998)}]{Ritus:1998jm}%
  \BibitemOpen
  \bibfield  {author} {\bibinfo {author} {\bibfnamefont {V.~I.}\ \bibnamefont
  {Ritus}},\ }\bibfield  {title} {\enquote {\bibinfo {title} {{Effective
  Lagrange function of intense electromagnetic field in QED}},}\ }in\
  \href@noop {} {\emph {\bibinfo {booktitle} {{Frontier tests of QED and
  physics of the vacuum. Proceedings, Workshop, Sandansky, Bulgaria, June 9-15,
  1998}}}}\ (\bibinfo {year} {1998})\ pp.\ \bibinfo {pages} {11--28},\ \Eprint
  {http://arxiv.org/abs/hep-th/9812124} {arXiv:hep-th/9812124 [hep-th]}
  \BibitemShut {NoStop}%
\bibitem [{\citenamefont {Dunne}(2004)}]{Dunne:2004nc}%
  \BibitemOpen
  \bibfield  {author} {\bibinfo {author} {\bibfnamefont {Gerald~V.}\
  \bibnamefont {Dunne}},\ }\bibfield  {title} {\enquote {\bibinfo {title}
  {{Heisenberg-Euler effective Lagrangians: Basics and extensions}},}\ }in\
  \href {\doibase 10.1142/9789812775344_0014} {\emph {\bibinfo {booktitle}
  {From fields to strings: Circumnavigating theoretical physics. Ian Kogan
  memorial collection (3 volume set)}}},\ \bibinfo {editor} {edited by\
  \bibinfo {editor} {\bibfnamefont {M.}~\bibnamefont {Shifman}}, \bibinfo
  {editor} {\bibfnamefont {A.}~\bibnamefont {Vainshtein}}, \ and\ \bibinfo
  {editor} {\bibfnamefont {J.}~\bibnamefont {Wheater}}}\ (\bibinfo {year}
  {2004})\ pp.\ \bibinfo {pages} {445--522},\ \Eprint
  {http://arxiv.org/abs/hep-th/0406216} {arXiv:hep-th/0406216 [hep-th]}
  \BibitemShut {NoStop}%
\bibitem [{\citenamefont {Gies}\ and\ \citenamefont
  {Karbstein}(2017)}]{Gies:2016yaa}%
  \BibitemOpen
  \bibfield  {author} {\bibinfo {author} {\bibfnamefont {Holger}\ \bibnamefont
  {Gies}}\ and\ \bibinfo {author} {\bibfnamefont {Felix}\ \bibnamefont
  {Karbstein}},\ }\bibfield  {title} {\enquote {\bibinfo {title} {{An Addendum
  to the Heisenberg-Euler effective action beyond one loop}},}\ }\href
  {\doibase 10.1007/JHEP03(2017)108} {\bibfield  {journal} {\bibinfo  {journal}
  {JHEP}\ }\textbf {\bibinfo {volume} {03}},\ \bibinfo {pages} {108} (\bibinfo
  {year} {2017})},\ \Eprint {http://arxiv.org/abs/1612.07251} {arXiv:1612.07251
  [hep-th]} \BibitemShut {NoStop}%
\bibitem [{\citenamefont {Huet}\ \emph {et~al.}(2017)\citenamefont {Huet},
  \citenamefont {Rausch~de Traubenberg},\ and\ \citenamefont
  {Schubert}}]{Huet:2017ydx}%
  \BibitemOpen
  \bibfield  {author} {\bibinfo {author} {\bibfnamefont {Idrish}\ \bibnamefont
  {Huet}}, \bibinfo {author} {\bibfnamefont {Michel}\ \bibnamefont {Rausch~de
  Traubenberg}}, \ and\ \bibinfo {author} {\bibfnamefont {Christian}\
  \bibnamefont {Schubert}},\ }\bibfield  {title} {\enquote {\bibinfo {title}
  {{Asymptotic behavior of the QED perturbation series}},}\ }\bibfield
  {booktitle} {\emph {\bibinfo {booktitle} {{5th Winter Workshop on
  Non-Perturbative Quantum Field Theory (WWNPQFT) Sophia-Antipolis, France,
  March 22-24, 2017}}},\ }\href {\doibase 10.1155/2017/6214341} {\bibfield
  {journal} {\bibinfo  {journal} {Adv. High Energy Phys.}\ }\textbf {\bibinfo
  {volume} {2017}},\ \bibinfo {pages} {6214341} (\bibinfo {year} {2017})},\
  \Eprint {http://arxiv.org/abs/1707.07655} {arXiv:1707.07655 [hep-th]}
  \BibitemShut {NoStop}%
\bibitem [{\citenamefont {Karbstein}(2019)}]{Karbstein:2019oej}%
  \BibitemOpen
  \bibfield  {author} {\bibinfo {author} {\bibfnamefont {Felix}\ \bibnamefont
  {Karbstein}},\ }\bibfield  {title} {\enquote {\bibinfo {title} {{Probing
  vacuum polarization effects with high-intensity lasers}},}\ \ }(\bibinfo
  {year} {2019})\ \Eprint {http://arxiv.org/abs/1912.11698} {arXiv:1912.11698
  [hep-ph]} \BibitemShut {NoStop}%
\bibitem [{\citenamefont {Weisstein}()}]{AssociatedLaguerrePolynomial}%
  \BibitemOpen
  \bibfield  {author} {\bibinfo {author} {\bibfnamefont {Eric~W.}\ \bibnamefont
  {Weisstein}},\ }\bibfield  {title} {\enquote {\bibinfo {title} {Associated
  laguerre polynomial},}\ }\href
  {http://mathworld.wolfram.com/AssociatedLaguerrePolynomial.html} {\bibinfo
  {journal} {From MathWorld--A Wolfram Web Resource.
  http://mathworld.wolfram.com/AssociatedLaguerrePolynomial.html}\
  }\BibitemShut {NoStop}%
\end{thebibliography}%

\end{document}